\pdfoutput=1
\RequirePackage{fix-cm}
\documentclass[12pt,a4paper]{article}
\usepackage{tikz}
\usetikzlibrary{arrows.meta}  % Add this line for newer arrow tips

\usepackage{multirow}
\usepackage{siunitx}

\usepackage{makecell}

\usepackage{ifthen} % for conditional statements
\newboolean{pdflatex}
\setboolean{pdflatex}{true} % False for eps figures 

\newboolean{articletitles}
\setboolean{articletitles}{true} % False removes titles in references

\newboolean{uprightparticles}
\setboolean{uprightparticles}{false} %True for upright particle symbols

\def\paperauthors{LHCb collaboration} % Leave as is for PAPER, CONF and FIGURE
\def\paperasciititle{Improved measurements of CP violation in Bs2Jpsipipi decays} % Set ASCII title here !! MAKE sure it's only ASCII characters !! 
\def\papertitle{Improved measurement of $C\!P$ violation in $B^{0}_{s} \!\to J/\psi \pi^{+}\pi^{-}$ decays} % Latex formatted title
\def\paperkeywords{{High Energy Physics}, {LHCb}} % Comma separated list
\def\papercopyright{\the\year\ CERN for the benefit of the LHCb collaboration} % new since 9/Apr/2018
\def\paperlicence{CC BY 4.0 licence}
\def\paperlicenceurl{https://creativecommons.org/licenses/by/4.0/}

\newif\ifEnableSectionTOCLinks
\EnableSectionTOCLinksfalse % deactivated

\usepackage[top=1in, bottom=1.25in, left=1in, right=1in]{geometry}

\usepackage{lineno}  % for line numbering during review
\usepackage{xspace} % To avoid problems with missing or double spaces after
\usepackage{caption} %these three command get the figure and table captions automatically small

\usepackage{graphicx}  % to include figures (can also use other packages)
\usepackage{color}
\usepackage{colortbl}
\graphicspath{{./figs/}} % Make Latex search fig subdir for figures
\usepackage{tikz}
\usetikzlibrary{angles}

\usepackage{amsmath} % Adds a large collection of math symbols
\usepackage{amssymb}
\usepackage{amsfonts}
\usepackage{upgreek} % Adds in support for greek letters in roman typeset

\newcommand*\patchAmsMathEnvironmentForLineno[1]{%
\expandafter\let\csname old#1\expandafter\endcsname\csname #1\endcsname
\expandafter\let\csname oldend#1\expandafter\endcsname\csname
end#1\endcsname
 \renewenvironment{#1}%
   {\linenomath\csname old#1\endcsname}%
   {\csname oldend#1\endcsname\endlinenomath}%
}
\newcommand*\patchBothAmsMathEnvironmentsForLineno[1]{%
  \patchAmsMathEnvironmentForLineno{#1}%
  \patchAmsMathEnvironmentForLineno{#1*}%
}
\AtBeginDocument{%
\patchBothAmsMathEnvironmentsForLineno{equation}%
\patchBothAmsMathEnvironmentsForLineno{align}%
\patchBothAmsMathEnvironmentsForLineno{flalign}%
\patchBothAmsMathEnvironmentsForLineno{alignat}%
\patchBothAmsMathEnvironmentsForLineno{gather}%
\patchBothAmsMathEnvironmentsForLineno{multline}%
\patchBothAmsMathEnvironmentsForLineno{eqnarray}%
}

\usepackage[pdftex,
            pdfauthor={\paperauthors},
            pdftitle={\paperasciititle},
            pdfkeywords={\paperkeywords}]{hyperref}
\usepackage{hyperxmp}
\hypersetup{
    pdfcopyright={Copyright (C) \papercopyright},
    pdflicenseurl={\paperlicenceurl}
}
\usepackage[colorinlistoftodos,textsize=scriptsize]{todonotes}

\usepackage[bottom,flushmargin,hang,multiple]{footmisc}

\usepackage[all]{hypcap} % Internal hyperlinks to floats.

\usepackage{xspace} 
\usepackage{upgreek}

\def\lhcb   {\mbox{LHCb}\xspace}

\def\bes    {\mbox{BES}\xspace}

\def\obelix {\mbox{OBELIX}\xspace}
\def\MagUp {\mbox{\em Mag\kern -0.05em Up}\xspace}

\ifthenelse{\boolean{uprightparticles}}%
{
 
 \def\Pgamma      {\ensuremath{\upgamma}\xspace}

 \def\Peta        {\ensuremath{\upeta}\xspace}

 \def\Pmu         {\ensuremath{\upmu}\xspace}

 \def\Ppi         {\ensuremath{\uppi}\xspace}                 
                  
 \def\Prho        {\ensuremath{\uprho}\xspace}

 \def\Pphi        {\ensuremath{\upphi}\xspace}

 \def\Ppsi        {\ensuremath{\uppsi}\xspace}

 \def\PDelta      {\ensuremath{\Delta}\xspace}                 
 \def\PXi         {\ensuremath{\Xi}\xspace}                 
 \def\PLambda     {\ensuremath{\Lambda}\xspace}                 
 \def\PSigma      {\ensuremath{\Sigma}\xspace}                 
 \def\POmega      {\ensuremath{\Omega}\xspace}                 
 \def\PUpsilon    {\ensuremath{\Upsilon}\xspace}
 \let\oldPi\Pi
 \def\PPi         {\ensuremath{\oldPi}\xspace}

 \def\PB      {\ensuremath{\mathrm{B}}\xspace}                 
 \def\PD      {\ensuremath{\mathrm{D}}\xspace}                 
 \def\PJ      {\ensuremath{\mathrm{J}}\xspace}                 
 \def\PK      {\ensuremath{\mathrm{K}}\xspace}                 
 \def\Pb      {\ensuremath{\mathrm{b}}\xspace}                 
 \def\Pc      {\ensuremath{\mathrm{c}}\xspace}                 
                  
 \def\Pe      {\ensuremath{\mathrm{e}}\xspace}                 
 \def\Pp      {\ensuremath{\mathrm{p}}\xspace}                 

 \def\Ps      {\ensuremath{\mathrm{s}}\xspace}                 
 \def\Pt      {\ensuremath{\mathrm{t}}\xspace}                 

 \def\thebaroffset{0.0em}
}
{
 
 \def\Pgamma      {\ensuremath{\gamma}\xspace}

 \def\Peta        {\ensuremath{\eta}\xspace}

 \def\Pmu         {\ensuremath{\mu}\xspace}

 \def\Ppi         {\ensuremath{\pi}\xspace}                 
                  
 \def\Prho        {\ensuremath{\rho}\xspace}

 \def\Pphi        {\ensuremath{\phi}\xspace}

 \def\Ppsi        {\ensuremath{\psi}\xspace}                 
                  
 \mathchardef\PDelta="7101
 \mathchardef\PXi="7104
 \mathchardef\PLambda="7103
 \mathchardef\PSigma="7106
 \mathchardef\POmega="710A
 \mathchardef\PUpsilon="7107
 \mathchardef\PPi="7105
 \def\PB      {\ensuremath{B}\xspace}                 
 \def\PD      {\ensuremath{D}\xspace}                 
 \def\PJ      {\ensuremath{J}\xspace}                 
 \def\PK      {\ensuremath{K}\xspace}                 
 \def\Pb      {\ensuremath{b}\xspace}                 
 \def\Pc      {\ensuremath{c}\xspace}                 
                  
 \def\Pe      {\ensuremath{e}\xspace}                 
 \def\Pp      {\ensuremath{p}\xspace}                 

 \def\Ps      {\ensuremath{s}\xspace}                 
 \def\Pt      {\ensuremath{t}\xspace}                 

 \def\thebaroffset{0.18em}
}
\newcommand{\offsetoverline}[2][\thebaroffset]{\kern #1\overline{\kern -#1 #2}}%

\makeatletter
\ifcase \@ptsize \relax% 10pt
  \newcommand{\miniscule}{\@setfontsize\miniscule{4}{5}}% \tiny: 5/6
\or% 11pt
  \newcommand{\miniscule}{\@setfontsize\miniscule{5}{6}}% \tiny: 6/7
\or% 12pt
  \newcommand{\miniscule}{\@setfontsize\miniscule{5}{6}}% \tiny: 6/7
\fi
\makeatother

\DeclareRobustCommand{\optbar}[1]{\shortstack{{\miniscule (\rule[.5ex]{1.25em}{.18mm})}
  \\ [-.7ex] $#1$}}

\def\en         {{\ensuremath{\Pe^-}}\xspace}   % electron negative (\em is taken)
\def\ep         {{\ensuremath{\Pe^+}}\xspace}

\def\mup        {{\ensuremath{\Pmu^+}}\xspace}
\def\mun        {{\ensuremath{\Pmu^-}}\xspace} % muon negative (\mum is taken)

\def\g      {{\ensuremath{\Pgamma}}\xspace}

\def\squark    {{\ensuremath{\Ps}}\xspace}

\def\cquark    {{\ensuremath{\Pc}}\xspace}
\def\cquarkbar {{\ensuremath{\overline \cquark}}\xspace}
\def\ccbar     {{\ensuremath{\cquark\cquarkbar}}\xspace}
\def\bquark    {{\ensuremath{\Pb}}\xspace}

\def\tquark    {{\ensuremath{\Pt}}\xspace}

\def\pion   {{\ensuremath{\Ppi}}\xspace}

\def\pip    {{\ensuremath{\pion^+}}\xspace}
\def\pim    {{\ensuremath{\pion^-}}\xspace}
\def\pipm   {{\ensuremath{\pion^\pm}}\xspace}

\def\rhomeson {{\ensuremath{\Prho}}\xspace}
\def\rhoz     {{\ensuremath{\rhomeson^0}}\xspace}

\def\kaon    {{\ensuremath{\PK}}\xspace}
\def\KorKbar {\kern \thebaroffset\optbar{\kern -\thebaroffset \PK}{}\xspace}

\def\Kp      {{\ensuremath{\kaon^+}}\xspace}
\def\Km      {{\ensuremath{\kaon^-}}\xspace}

\def\Kstarz  {{\ensuremath{\kaon^{*0}}}\xspace}

\newcommand{\etapr}{\ensuremath{\Peta^{\prime}}\xspace}
\newcommand{\phiz}{\ensuremath{\Pphi}\xspace}

\def\D       {{\ensuremath{\PD}}\xspace}

\def\DorDbar {\kern \thebaroffset\optbar{\kern -\thebaroffset \PD}\xspace}

\def\Dp      {{\ensuremath{\D^+}}\xspace}
\def\Dm      {{\ensuremath{\D^-}}\xspace}

\def\DpDm    {\ensuremath{\Dp {\kern -0.16em \Dm}}\xspace}

\def\Dsp     {{\ensuremath{\D^+_\squark}}\xspace}
\def\Dsm     {{\ensuremath{\D^-_\squark}}\xspace}

\def\B       {{\ensuremath{\PB}}\xspace}
\def\Bbar    {{\ensuremath{\offsetoverline{\PB}}}\xspace}

\def\BorBbar {\kern \thebaroffset\optbar{\kern -\thebaroffset \PB}\xspace}

\def\Bd      {{\ensuremath{\B^0}}\xspace}

\def\BdorBdbar {\ensuremath{\kern \thebaroffset\optbar{\kern -\thebaroffset \PB}{}^0}\xspace}
\def\Bu      {{\ensuremath{\B^+}}\xspace}

\def\Bs      {{\ensuremath{\B^0_\squark}}\xspace}
\def\Bsb     {{\ensuremath{\Bbar{}^0_\squark}}\xspace}
\def\BsorBsbar {\ensuremath{\kern \thebaroffset\optbar{\kern -\thebaroffset \PB}{}^0_{\!\!\!\squark}}\xspace}
\def\Bc      {{\ensuremath{\B_\cquark^+}}\xspace}

\def\Bds     {{\ensuremath{\B_{(\squark)}^0}}\xspace}

\def\jpsi     {{\ensuremath{{\PJ\mskip -3mu/\mskip -2mu\Ppsi}}}\xspace}
\def\psitwos  {{\ensuremath{\Ppsi{(2S)}}}\xspace}

\def\Y#1S{\ensuremath{\PUpsilon{(#1S)}}\xspace}

\def\proton      {{\ensuremath{\Pp}}\xspace}
\def\antiproton  {{\ensuremath{\overline \proton}}\xspace}

\def\Lz          {{\ensuremath{\PLambda}}\xspace}

\def\LorLbar     {\kern \thebaroffset\optbar{\kern -\thebaroffset \PLambda}\xspace}

\def\Lb           {{\ensuremath{\Lz^0_\bquark}}\xspace}

\def\BF         {{\ensuremath{\mathcal{B}}}\xspace}

\newcommand{\decay}[2]{\ensuremath{\mathinner{#1\!\to #2}}\xspace}

\def\to                 {\ensuremath{\rightarrow}\xspace}

\newcommand{\tauL}{{\ensuremath{\tau_{\mathrm{ L}}}}\xspace}
\newcommand{\tauH}{{\ensuremath{\tau_{\mathrm{ H}}}}\xspace}

\def\CP                {{\ensuremath{C\!P}}\xspace}

\def\Vij  {{\ensuremath{V_{ij}^{\phantom{\ast}}}}\xspace}

\def\Vcs  {{\ensuremath{V_{\cquark\squark}^{\phantom{\ast}}}}\xspace}

\def\Vtb  {{\ensuremath{V_{\tquark\bquark}^{\phantom{\ast}}}}\xspace}

\def\Vtss  {{\ensuremath{V_{\tquark\squark}^\ast}}\xspace}

\def\Vcbs  {{\ensuremath{V_{\cquark\bquark}^\ast}}\xspace}

\newcommand{\dms}{{\ensuremath{\Delta m_{\squark}}}\xspace}

\newcommand{\cdms}{{\ensuremath{\cos(\dms t)}}\xspace}

\newcommand{\sdms}{{\ensuremath{\sin(\dms t)}}\xspace}

\newcommand{\mL}{{\ensuremath{m_{\mathrm{ L}}}}\xspace}
\newcommand{\mH}{{\ensuremath{m_{\mathrm{ H}}}}\xspace}

\newcommand{\DGs}{{\ensuremath{\Delta\Gamma_{\squark}}}\xspace}

\newcommand{\Gs}{{\ensuremath{\Gamma_{\squark}}}\xspace}

\newcommand{\GorGb}{\kern \thebaroffset\optbar{\kern -\thebaroffset \Gamma}\xspace}
\newcommand{\cDGs}{\cosh\!\left(\frac{\DGs}{2}t\right)}

\newcommand{\sDGs}{\sinh\!\left(\frac{\DGs}{2}t\right)}

\newcommand{\GL}{{\ensuremath{\Gamma_{\mathrm{ L}}}}\xspace}
\newcommand{\GH}{{\ensuremath{\Gamma_{\mathrm{ H}}}}\xspace}

\newcommand{\phis}{{\ensuremath{\phi_{\squark}}}\xspace}

\def\AT#1     {\ensuremath{A_{\mathrm{T}}^{#1}}\xspace}           % 2

\def\C#1      {\ensuremath{\mathcal{C}_{#1}}\xspace}                       % 9
\def\Cp#1     {\ensuremath{\mathcal{C}_{#1}^{'}}\xspace}                    % 7
\def\Ceff#1   {\ensuremath{\mathcal{C}_{#1}^{\mathrm{(eff)}}}\xspace}        % 9  
\def\Cpeff#1  {\ensuremath{\mathcal{C}_{#1}^{'\mathrm{(eff)}}}\xspace}       % 7
\def\Ope#1    {\ensuremath{\mathcal{O}_{#1}}\xspace}                       % 2
\def\Opep#1   {\ensuremath{\mathcal{O}_{#1}^{'}}\xspace}                    % 7

\newcommand{\nospaceunit}[1]{\ensuremath{\text{#1}}}       
\newcommand{\aunit}[1]{\ensuremath{\text{\,#1}}}       
\newcommand{\tev}{\aunit{Te\kern -0.1em V}\xspace}
\newcommand{\gev}{\aunit{Ge\kern -0.1em V}\xspace}
\newcommand{\mev}{\aunit{Me\kern -0.1em V}\xspace}
\newcommand{\kev}{\aunit{ke\kern -0.1em V}\xspace}
\newcommand{\ev}{\aunit{e\kern -0.1em V}\xspace}
 
\newcommand{\mevc}{\ensuremath{\aunit{Me\kern -0.1em V\!/}c}\xspace}
\newcommand{\gevc}{\ensuremath{\aunit{Ge\kern -0.1em V\!/}c}\xspace}
\newcommand{\mevcc}{\ensuremath{\aunit{Me\kern -0.1em V\!/}c^2}\xspace}
\newcommand{\gevcc}{\ensuremath{\aunit{Ge\kern -0.1em V\!/}c^2}\xspace}
\def\mum  {\ensuremath{\,\upmu\nospaceunit{m}}\xspace}

\def\fb   {\ensuremath{\aunit{fb}}\xspace}
\def\invfb   {\ensuremath{\fb^{-1}}\xspace}

\def\ps   {\ensuremath{\aunit{ps}}\xspace}
\def\fs   {\ensuremath{\aunit{fs}}\xspace}

\def\invfs{\ensuremath{\fs^{-1}}\xspace}
\def\invps{\ensuremath{\ps^{-1}}\xspace}

\newcommand{\chisq}{\ensuremath{\chi^2}\xspace}
\newcommand{\chisqndf}{\ensuremath{\chi^2/\mathrm{ndf}}\xspace}
\newcommand{\chisqip}{\ensuremath{\chi^2_{\text{IP}}}\xspace}

\def\gsim{{~\raise.15em\hbox{$>$}\kern-.85em
          \lower.35em\hbox{$\sim$}~}\xspace}
\def\lsim{{~\raise.15em\hbox{$<$}\kern-.85em
          \lower.35em\hbox{$\sim$}~}\xspace}

\newcommand{\abs}[1]{\ensuremath{\left|#1\right|}} % {x}
\newcommand{\mbracket}[1]{\ensuremath{\left(#1\right)}} % (x)
\newcommand{\cbracket}[1]{\ensuremath{\left[#1\right]}} % [x]
\newcommand{\Real}{\ensuremath{\mathcal{R}e}\xspace}
\newcommand{\Imag}{\ensuremath{\mathcal{I}m}\xspace}

\newcommand{\tabincell}[2]{\begin{tabular}{@{}#1@{}}#2\end{tabular}}

\def\themu   {\ensuremath{\mathrm{\theta_{\mu}}}\xspace}
\def\thepi   {\ensuremath{\mathrm{\theta_{\pi}}}\xspace}

\def\cosmu   {\ensuremath{\mathrm{\cos\themu}}\xspace}
\def\cospi   {\ensuremath{\mathrm{\cos\thepi}}\xspace}

\newcommand{\mkk}{\ensuremath{m_{KK}}\xspace}
\newcommand{\mpipi}{\ensuremath{m_{\pi\pi}}\xspace}

\newcommand{\tprime}{{\ensuremath{t^{\prime}}}\xspace}

\newcommand{\pdfp}{{\ensuremath{\mathcal{P}}}\xspace}

\def\sqs   {\ensuremath{\protect\sqrt{s}}\xspace}

\def\pt         {\ensuremath{p_{\mathrm{T}}}\xspace}

\def\ptot       {\ensuremath{p}\xspace}

\def\mrad{\aunit{mrad}\xspace}
\def\rad{\aunit{rad}\xspace}

\def\evtgen     {\mbox{\textsc{EvtGen}}\xspace}

\def\geant      {\mbox{\textsc{Geant4}}\xspace}

\def\photos     {\mbox{\textsc{Photos}}\xspace}

\def\pythia     {\mbox{\textsc{Pythia}}\xspace}

\def\tell1  {TELL1\xspace}
\def\ukl1   {UKL1\xspace}

\def\runone {\mbox{Run 1}\xspace}
\def\runtwo {\mbox{Run 2}\xspace}

\newcommand{\ie}{\mbox{\itshape i.e.}\xspace}

\newcommand{\lhcborcid}[1]{\href{https://orcid.org/#1}{\hspace*{0.1em}\raisebox{-0.45ex}{\includegraphics[width=1em]{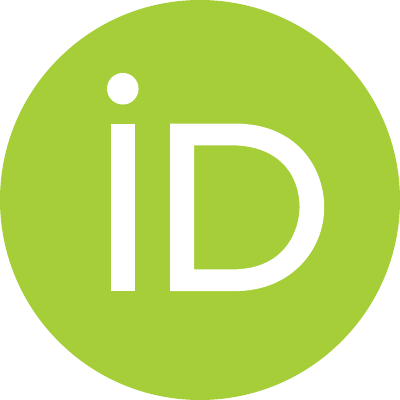}}}}

\hypersetup{
  colorlinks   = true, %Colours links instead of ugly boxes
  urlcolor     = blue, %Colour for external hyperlinks
  linkcolor    = blue, %Colour of internal links
  citecolor    = red   %Colour of citations
}

\ifEnableSectionTOCLinks
    \usepackage[explicit]{titlesec} % to change headings
    
    \let\oldcontentsline\contentsline
    \renewcommand

    \titleformat{\section}{\normalfont\Large\bf}{\hyperlink{tocsection.\thesection}{{\thesection} \parbox[t]{\dimexpr\textwidth-1pc}{#1}}}{1pc}{}

    \titleformat{\subsection}{\normalfont\bf}{\hyperlink{tocsubsection.\thesubsection}{{\thesubsection} \parbox[t]{\dimexpr\textwidth-1pc}{#1}}}{1pc}{}

    \titleformat{name=\section,numberless}[display]{}{}{0pt}{\normalfont\Huge\bfseries #1}
\fi

\usepackage{cite} % Allows for ranges in citations
\usepackage{mciteplus}
\usepackage{longtable} % only for template; not usually to be used in PAPERs
\usepackage{placeins} % allows use of \FloatBarrier to control float placement

\begin{document}

%%%%%%%%%%%%%%%%%%%%%%%%%
%%%%% Title     %%%%%%%%%
%%%%%%%%%%%%%%%%%%%%%%%%%
\renewcommand{\thefootnote}{\fnsymbol{footnote}}
\setcounter{footnote}{1}

% %%%%%%% CHOOSE TITLE PAGE--------
%\onecolumn
%\input{title-LHCb-INT}
%\input{title-LHCb-ANA}
%\input{title-LHCb-CONF}
%\input{title-LHCb-FIGURE}
% ===============================================================================
% Purpose: LHCb-PAPER journal paper title page template
% Author: 
% Created on: 2010-09-25
% ===============================================================================

%%%%%%%%%%%%%%%%%%%%%%%%%
%%%%%  TITLE PAGE  %%%%%%
%%%%%%%%%%%%%%%%%%%%%%%%%
\begin{titlepage}
\pagenumbering{roman}

% Header ---------------------------------------------------
\vspace*{-1.5cm}
\centerline{\large EUROPEAN ORGANIZATION FOR NUCLEAR RESEARCH (CERN)}
\vspace*{1.5cm}
\noindent
\begin{tabular*}{\linewidth}{lc@{\extracolsep{\fill}}r@{\extracolsep{0pt}}}
\ifthenelse{\boolean{pdflatex}}% Logo format choice
{\vspace*{-1.5cm}\mbox{\!\!\!\includegraphics[width=.14\textwidth]{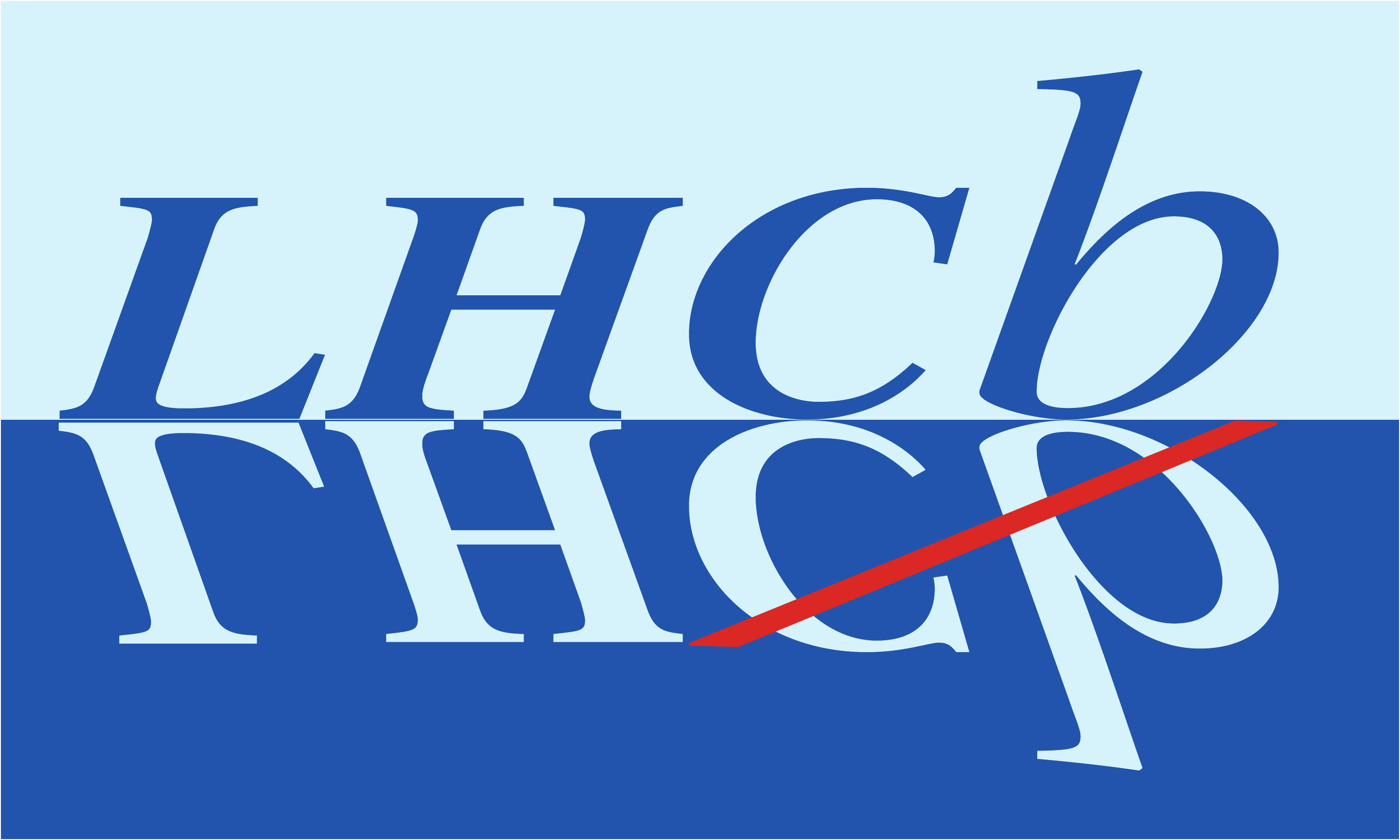}} & &}%
{\vspace*{-1.2cm}\mbox{\!\!\!\includegraphics[width=.12\textwidth]{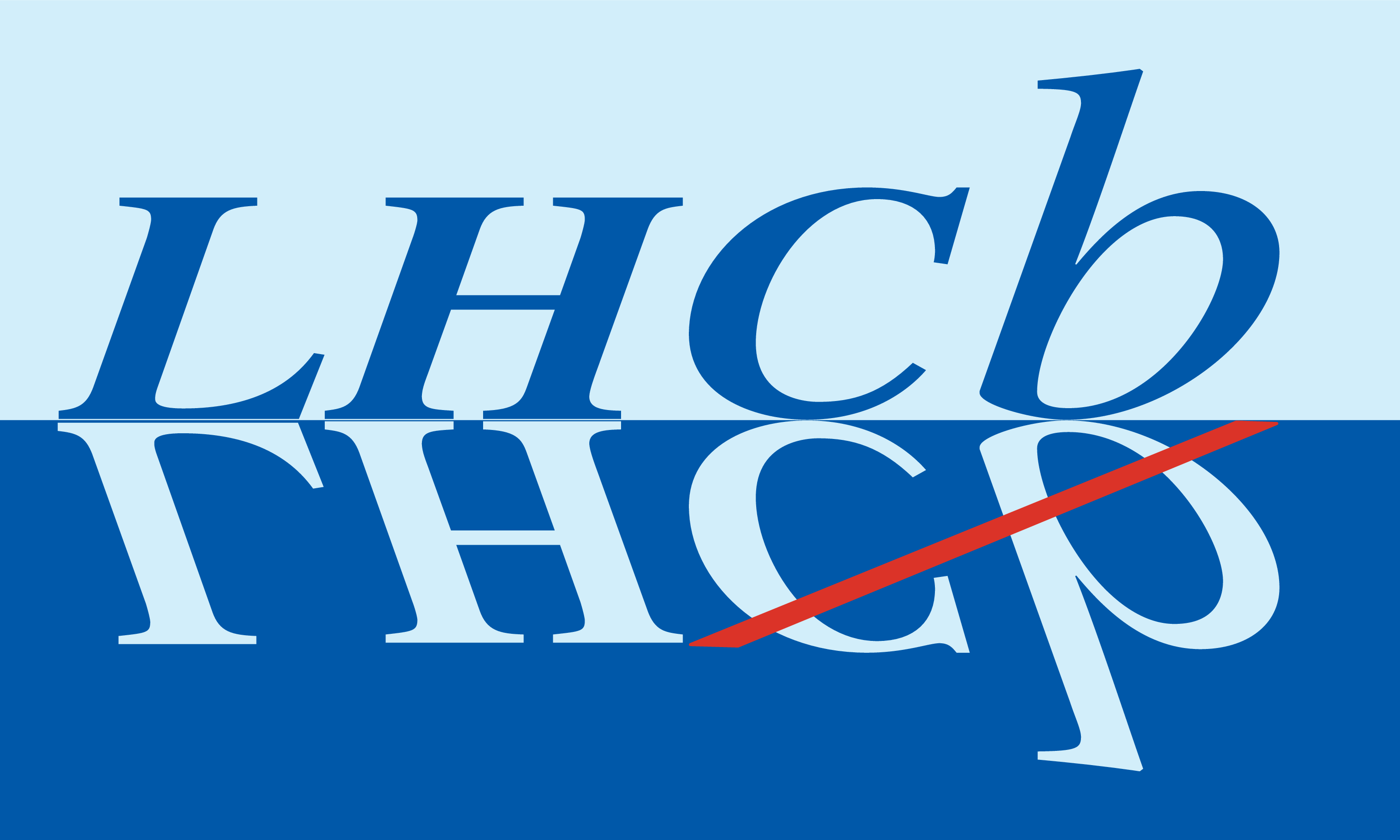}} & &}%
\\
 & & CERN-EP-2026-181 \\  % ID 
 & & LHCb-PAPER-2026-017 \\  % ID 
 & & 14 August 2026 \\ % Date - Can also hardwire e.g.: 23 March 2010
 & & \\
% not in paper \hline
\end{tabular*}

\vspace*{4.0cm}

% Title --------------------------------------------------
{\normalfont\bfseries\boldmath\huge
\begin{center}
% DO NOT EDIT HERE. Instead edit macro in main.tex to keep metadata correct
  \papertitle 
\end{center}
}

\vspace*{1.5cm}

% Authors -------------------------------------------------
\begin{center}
%In the footnote, replace 'paper' by 'Letter' in case of submission to PRL or PLB 
% Edit macro in main.tex to keep metadata correct
\paperauthors\footnote{Authors are listed at the end of this paper.}
\end{center}

%\vspace{\fill}

% Abstract -----------------------------------------------
\begin{abstract}
  \noindent
 
  The time-dependent $C\!P$ asymmetry in $B^{0}_{s} \!\to J/\psi \pi^{+}\pi^{-}$ decays is measured using proton-proton collision data, corresponding to an integrated luminosity of~$6\ensuremath{\aunit{fb}\xspace^{-1}}\xspace$, 
  collected with the LHCb detector at a centre-of-mass energy of $13\aunit{Te\kern -0.1em V}\xspace$ during $\mbox{2015--2018}$. 
  The $C\!P$-violating phase, $\phi_{s}$, the direct $C\!P$-violation parameter, $\left|\lambda\right|$, 
  and the decay width of the heavy mass eigenstate in the $B^{0}_{s}$ system, $\Gamma_{\mathrm{ H}}$,
  are measured respectively to be 
  $\phi_{s} = -0.077 \pm 0.034 \pm 0.007\ensuremath{\aunit{rad}}\xspace$, 
  $\left|\lambda\right| = 0.993 \pm 0.026 \pm 0.007$
  and $\Gamma_{\mathrm{ H}} = 0.610 \pm 0.002 \pm 0.004\ensuremath{\aunit{ps}^{-1}}\xspace$,
  where the first uncertainties are statistical
  and the second systematic.
  These results are consistent with previous measurements and the expectation based on the Standard Model.
  The combination with previous measurements in $B^{0}_{s} \!\to J/\psi \pi^{+}\pi^{-}$ decays using $7\aunit{Te\kern -0.1em V}\xspace$ and $8\aunit{Te\kern -0.1em V}\xspace$ proton-proton collision data yields 
  $\phi_{s} = -0.046 \pm 0.031\ensuremath{\aunit{rad}}\xspace$,
  $\left|\lambda\right| = 0.975 \pm 0.024$
  and $\Gamma_{\mathrm{ H}} = 0.610 \pm 0.004\ensuremath{\aunit{ps}^{-1}}\xspace$,
  while the combination including all other LHCb measurements gives
  $\phi_{s} = -0.041 \pm 0.017\ensuremath{\aunit{rad}}\xspace$.

\end{abstract}

\vspace*{0.5cm}

\begin{center}
%  To be submitted to
  Submitted to
  JHEP 
%  Phys.~Rev.~D 
%  Phys.~Rev.~Lett. /
%  Phys.~Lett.~B /
%  Eur.~Phys.~J.~C /
%  Nucl.~Phys.~B /
%  Chin.~Phys.~C /
%  Nature~Physics /
%  sciPost~Physics /
%  J. Instr. /
%  Instruments 
\end{center}

\vspace{\fill}

{\footnotesize 
% Edit macro in main.tex to keep metadata correct
\centerline{\copyright~\papercopyright. \href{\paperlicenceurl}{\paperlicence}.}}
\vspace*{2mm}

\end{titlepage}

%%%%%%%%%%%%%%%%%%%%%%%%%%%%%%%%
%%%%%  EOD OF TITLE PAGE  %%%%%%
%%%%%%%%%%%%%%%%%%%%%%%%%%%%%%%%

%  empty page follows the title page ----
\newpage
\setcounter{page}{2}
\mbox{~}
%\newpage
%
%% Author List ----------------------------
%%  You need to get a new author list!
%\input{Authorship_LHCb-PAPER-2026-017}
%
%The author list for journal publications is provided by the Membership Committee shortly after 'approval to go to paper' has been given.
%%It will be made available on the page
%%\verb!http://www.physik.uzh.ch/~strauman/forMemCo/LHCb-PAPER-XXXX-XXX/! .
%It will be sent to you by email shortly after a paper number has beens assigned.
%The author list should be included already at first circulation, 
%to allow new members of the collaboration to verify whether they have been included correctly.
%Occasionally a misspelled name is corrected or associated institutions become full members.
%In that case, a new author list will be sent to you.
%In case line numbering doesn't work well after including the authorlist, try moving the \verb!\bigskip! after the last author to a separate line.
%
%
%The authorship for Conference Reports should be ``The LHCb
%  collaboration'', with a footnote giving the name(s) of the contact
%  author(s), but without the full list of collaboration names.

%\twocolumn
% %%%%%%%%%%%%% ---------

\renewcommand{\thefootnote}{\arabic{footnote}}
\setcounter{footnote}{0}

%%%%%%%%%%%%%%%%%%%%%%%%%%%%%%%%
%%%%%  Table of Content   %%%%%%
%%%%%%%%%%%%%%%%%%%%%%%%%%%%%%%%
%%%% Uncomment if desired
%\tableofcontents

\cleardoublepage

%%%%%%%%%%%%%%%%%%%%%%%%%
%%%%% Main text %%%%%%%%%
%%%%%%%%%%%%%%%%%%%%%%%%%

\pagestyle{plain} % restore page numbers for the main text
\setcounter{page}{1}
\pagenumbering{arabic}

%% Uncomment during review phase. 
%% Comment before a final submission.
%\linenumbers

%% This is the main body
%% It is useful to have a single file so comments are not missed in overleaf.
\section{Introduction}
\label{sec:Introduction}
 
The \CP violation in the $\Bs$ meson system,
arising from the interference of the amplitude of the direct decay and that of the adjoint decay preceded by $\Bs$-$\Bsb$ oscillation,
is parameterised by the \CP-violating phase, $\phis$.
In the Standard Model (SM), 
a global fit to experimental data determines that
$\phi_s^{\rm SM} \equiv -2\arg(-\frac{\Vtb \Vtss}{\Vcs \Vcbs}) = -0.0376^{+0.0006}_{-0.0005}\rad$~\cite{CKMfitter2015},
where $\Vij$ are elements of the Cabibbo--Kobayashi--Maskawa (CKM) matrix~\cite{Cabibbo:1963yz,*Kobayashi:1973fv}.
This precise prediction establishes the measurement of $\phis$ as an excellent test of the SM,
as well as a sensitive probe to the physics beyond the SM~\cite{Nir:1990hj}. 

Usually, the decays proceeding via the $\decay{\bquark}{\ccbar\squark}$ transition are employed to measure~$\phis$~\cite{Dunietz:2000cr}.
Several experiments have measured $\phis$
in the golden decay\footnote{The inclusion of charge-conjugate processes is implied throughout.} $\decay{\Bs}{\jpsi(\mup\mun)\phi}$
\cite{CDF:2012nqr,D0:2011ymu,ATLAS:2014nmm,ATLAS:2016pno,ATLAS:2020lbz,CMS:2015asi,CMS:2024znt,LHCb-PAPER-2014-059,LHCb-PAPER-2023-016}, yielding an average value of $\phi_s^{\jpsi\phi} = -0.060 \pm 0.014\rad$~\cite{HFLAV23-Zenodo-19446771}.
Benefiting from the excellent performance of the \lhcb detector,
this phase has also been measured in other decay modes,
including $\decay{\Bs}{\jpsi\Kp\Km}$ (with \mbox{$\mkk > 1.05\gevcc$})~\cite{LHCb-PAPER-2017-008},
$\decay{\Bs}{\jpsi(\ep\en)\phiz}$~\cite{LHCb-PAPER-2020-042},
$\decay{\Bs}{\psitwos\phiz}$~\cite{LHCb-PAPER-2016-027},
$\decay{\Bs}{\Dsp\Dsm}$~\cite{LHCb-PAPER-2014-051,LHCb-PAPER-2024-027},
and $\decay{\Bs}{\jpsi\pip\pim}$ decays~\cite{LHCb-PAPER-2014-019,LHCb-PAPER-2019-003}. 
All of the measurements neglect subleading electroweak loop (penguin) diagrams contributions.
The size of the penguin contributions has been constrained using $\decay{\Bd}{\jpsi\rhoz}$ decays~\cite{LHCb-PAPER-2025-059},
giving a phase shift of $\Delta\phis=5.0\pm4.6\mrad$,
which is consistent with zero.
Among these decays, the $\decay{\Bs}{\jpsi\pip\pim}$ decays provide the second-best sensitivity to the $\phis$ phase
and improve the precision of the \lhcb average of $\phi_s^{\ccbar\squark}$ by approximately $15\%$.
The combination of these measurements leads to a world average value of $\phi_s^{\ccbar\squark} = -0.052 \pm 0.013\rad$~\cite{HFLAV23-Zenodo-19446771}, 
whose precision is far from that of $\phi_s^{\rm SM}$.
Therefore, further improving the precision of $\phis$ is essential for a more stringent test of the SM. 

The phase $\phis$ has been measured in $\decay{\Bs}{\jpsi\pip\pim}$ decays
using proton-proton ($pp$) collision data collected by the \lhcb detector 
corresponding to an integrated luminosity of $3\invfb$ at centre-of-mass energies $\sqrt{s}=7$ and $8\tev$ in 2011--2012 (\runone)~\cite{LHCb-PAPER-2014-019}
and $2\invfb$ at $\sqrt{s}=13\tev$ in 2015--2016 (part of \runtwo)~\cite{LHCb-PAPER-2019-003}.
This paper reports an updated $\phis$ measurement in $\decay{\Bs}{\jpsi\pip\pim}$ decays with $\decay{\jpsi}{\mup\mun}$, using the full data collected during $\mbox{2015--2018}$ (\runtwo) at $\sqs = 13\tev$, corresponding to an integrated luminosity of $6\invfb$.
In this measurement,
subleading penguin contributions are also neglected.
To account for varying data-taking conditions, the data are split into three datasets, corresponding to the years 2015--2016, 
2017, and 2018, which are analysed independently but fitted simultaneously to extract the physics parameters of interest.

This paper is organised as follows.
Section~\ref{sec:Amplitude} introduces the time-dependent decay rates of $\decay{\Bs}{\jpsi\pip\pim}$. 
A brief description of the \lhcb detector, trigger, and simulation is presented in Sec.~\ref{sec:Detector}.
The event selection and signal extraction are detailed in Sec.~\ref{sec:Selection} and Sec.~\ref{Sec:Signal}, respectively.
Section~\ref{sec:DetectorEffects} covers the studies of detector effects, including decay-time resolution, 
decay-time efficiency and angular efficiency,
and Sec.~\ref{Sec:FlavTag} presents the flavour-tagging technique~\cite{Fazzini:2018dyq, LHCb-PAPER-2011-027,LHCb-PAPER-2015-027,LHCb-PAPER-2015-056} used to identify the flavour of the $\BsorBsbar$ meson at production,
which is essential for measurements of time-dependent \CP violation. 
A description of $\pip\pim$ resonance structure is given in Sec.~\ref{Sec:Resonances},
where the complex $\pip\pim$ spectrum from overlapping scalar and tensor resonances
is modelled to reliably extract $\phis$.
The measurement of the \CP-violation parameters is presented in Sec.~\ref{sec:Results},
followed by the study of systematic uncertainties in Sec.~\ref{sec:Systematics}.
Finally, a conclusion is given in Sec.~\ref{sec:Conclusion}.

\section{Phenomenology}
\label{sec:Amplitude}

The measurement of \CP violation in $\decay{\Bs}{\jpsi\pip\pim}$ decays requires a complete analysis of the angular decay structure,
decay time, and $\pip\pim$ invariant mass to separate different transversity components arising from the polarisations of $\jpsi$ and orbital angular momentum of the $\pip\pim$ system.
The time-dependent decay rates for $\decay{\Bs}{\jpsi\pip\pim}$ and $\decay{\Bsb}{\jpsi\pip\pim}$ decays, $\Gamma(t)$ and $\overline{\Gamma}(t)$,
are given as~\cite{LHCb-PAPER-2019-003}
\begin{eqnarray}
\label{eq:time-dependent-decay-rate}
\GorGb(t) \propto e^{-\Gs t} 
\left\{\frac{1}{2} (\abs{\cal A}^2 + \abs{\overline{\cal A}}^2) \cDGs  \pm 
 \frac{1}{2} (\abs{\cal A}^2 - \abs{\overline{\cal A}}^2) \cdms \right.\quad\quad\nonumber\\
- \left.\Real({\cal A}^\ast \overline{\cal A}) \sDGs  \mp  \Imag({\cal A}^\ast \overline{\cal A}) \sdms\right\}\;,
\end{eqnarray}
where $t$ is the decay time in the $\Bs$ rest frame,
$\Gs \equiv (\GL + \GH)/2$ is the average decay width,
$\DGs \equiv \GL - \GH$ is the decay width difference, 
$\dms \equiv \mH - \mL$ is the mass difference,
and $\mH$ ($\mL$) and $\GH$ ($\GL$) are the mass and decay width of the heavy (light) mass eigenstate~\cite{Nierste:2009wg}, respectively.
In Eq.~\ref{eq:time-dependent-decay-rate}, $\Bs$ and $\Bsb$ refer to the flavour at production,
and it is assumed that $\abs{q/p} = 1$,
which is confirmed by current measurements at the per-mille level~\cite{HFLAV23-Zenodo-19446771},
where $p$ and $q$ are the mixing parameters relating the $\Bs$ mass eigenstates to the strong interaction eigenstates.
The total decay amplitude, ${\cal A}$, is the sum
over the individual transversity amplitudes, ${\cal A}_i$, of the $\pip\pim$ resonant
and nonresonant states~\cite{Dighe:1995pd}.
These transversity amplitudes, labelled longitudinal (0), parallel ($\parallel$), and perpendicular ($\perp$),
reflect the polarisation of the $\jpsi$ meson and the orbital angular momentum of the $\pip\pim$ system.
The \CP-conjugate total amplitude is given by 
$\overline{\cal A} \equiv \sum_i \eta_i \abs{\lambda_i} e^{-i \phi_s^i} {\cal A}_i$,
where $\eta_i$ is the \CP eigenvalue, $\lambda_i$ is the direct \CP-violation 
parameter for each component, and 
$\phi_s^i \equiv -\arg(\eta_i \lambda_i)$ is the corresponding \CP-violating phase~\cite{LHCb-PAPER-2013-002}.
In this analysis, \CP violation is assumed to be independent of the transversity states, 
\ie $\lambda = \eta_i \lambda_i$
and $\phis = -\arg(\lambda)$ are the same for all components.

\begin{figure}[tb]
     \begin{center}
     \includegraphics[width=0.8\linewidth]{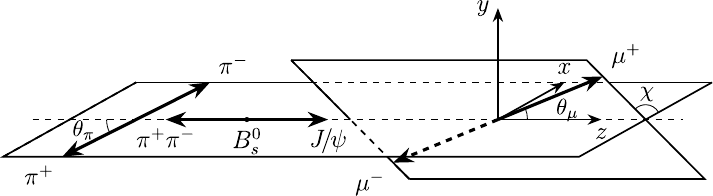}
     \end{center}
     \caption{
     Definition of the helicity angles, $\theta_\pi$, $\theta_\mu$ and $\chi$, in the $B^{0}_{s} \!\to J/\psi \pi^{+}\pi^{-}$ decays.
     }
     \label{Fig1}
\end{figure}

Previous analyses of the resonant structure
in the $\pip\pim$ mass spectrum~\cite{LHCb-PAPER-2012-005, LHCb-PAPER-2013-069, LHCb-PAPER-2019-003} of $\decay{\Bs}{\jpsi\pip\pim}$ decays
have identified contributions from scalar mesons, $f_0$, and tensor mesons, $f_{2}$.
These studies show that $\decay{\Bs}{\jpsi\pip\pim}$ decays are dominated by the \CP-odd component~\cite{LHCb-PAPER-2013-069},
which enables a direct measurement of the decay width of the heavy mass eigenstate $\GH$.
This analysis takes into account the \CP-even component for measuring $\phis$~\cite{Zhang:2012zk}.
The polarisation amplitude of each resonant component 
is a function of the $\pip\pim$ invariant mass, $\mpipi$,
and the decay angles, 
$\boldsymbol{\Omega} \equiv (\cospi,\,\cosmu,\,\chi)$,
which are shown in Fig.~\ref{Fig1}.
The helicity angle, $\thepi(\themu)$, is 
the angle between the $\pip(\mup)$ momentum direction in the $\pip\pim(\jpsi)$ rest frame and the $\pip\pim(\jpsi)$ momentum direction in the $\Bs$ rest frame,
and $\chi$ is the angle between the $\jpsi$ and $\pip\pim$ decay planes.\footnote{The definitions are the same for $\Bs$ and $\Bsb$ decays.}
The specific calculations of $\abs{{\cal A}}^2$,
$\abs{\overline{{\cal A}}}^2$,
and ${\cal A}^\ast \overline{\cal A}$
in terms of $\mpipi$ and $\boldsymbol{\Omega}$
are described in Ref.~\cite{Zhang:2012zk}.

%%%%%%%%%%%%%%%%%%%%%%%%%%%%%%%%%%%%%%%%%%%%%%%%
%%%%%%%%%%%%%%%%%%%%%%%%%%%%%%%%%%%%%%%%%%%%%%%%

The amplitude analysis is performed with an unbinned maximum-likelihood fit 
to the $\pip\pim$ invariant mass, decay angles, and decay-time distributions
of a background-subtracted sample of the $\decay{\Bs}{\jpsi\pip\pim}$ decays. 
The detector effects, including decay-time resolution, decay-time and angular efficiencies,
as well as flavour tagging~\cite{Fazzini:2018dyq, LHCb-PAPER-2011-027,LHCb-PAPER-2015-027,LHCb-PAPER-2015-056},
are accounted for in the fit in Sec.~\ref{sec:Results}.

\section{Detector, trigger and simulation}
\label{sec:Detector}

The \lhcb detector~\cite{LHCb-DP-2008-001, LHCb-DP-2014-002} is a single-arm forward
spectrometer covering the pseudorapidity range $2 < \eta < 5$,
designed for the study of particles containing $\bquark$ or $\cquark$
quarks. The detector used to collect the data analysed in this paper
includes a high-precision tracking system
consisting of a silicon-strip vertex detector surrounding the $pp$
interaction region~\cite{LHCb-DP-2014-001}, a large-area silicon-strip detector located
upstream of a dipole magnet with a bending power of about
$4{\mathrm{\,T\,m}}$, and three stations of silicon-strip detectors and straw
drift tubes~\cite{LHCb-DP-2017-001} 
placed downstream of the magnet.
The polarity of the magnetic field is regularly reversed in order to cancel detection asymmetries.
The tracking system provides a measurement of the momentum, $\ptot$, of charged particles with
a relative uncertainty that varies from 0.5\% at low momentum to $1.0\%$ at $200\gevc$.
The minimum distance of a track to a primary $pp$ collision vertex (PV), the impact parameter (IP),
is measured with a resolution of $(15+29/\pt)\mum$,
where $\pt$ is the component of the momentum transverse to the beam, in\,$\gevc$.
Different types of charged hadrons are distinguished using information
from two ring-imaging Cherenkov detectors~\cite{LHCb-DP-2012-003}.
Photons, electrons and hadrons are identified by a calorimeter system consisting of
scintillating-pad and preshower detectors, an electromagnetic
and a hadronic calorimeter. Muons are identified by a
system composed of alternating layers of iron and multiwire
proportional chambers~\cite{LHCb-DP-2012-002}.

The online event selection is performed by a trigger system~\cite{LHCb-DP-2012-004},
which consists of a hardware stage based on information from the calorimeter and muon
systems, and a software stage that applies a full event
reconstruction.
Two distinct categories of software triggers are employed to maximise the signal yield,
where the first requires a secondary vertex with a large IP significance relative to any PV,
whereas the second does not.
Therefore, the two trigger categories introduce different efficiencies as a function of the decay time, the $\pip\pim$ invariant mass, and the decay angles.

Simulation is required to model the effects of the detector efficiency and the
imposed selection requirements.
In the simulation, $pp$ collisions are generated using
\pythia~\cite{Sjostrand:2007gs, *Sjostrand:2006za} 
with a specific \lhcb configuration~\cite{LHCb-PROC-2010-056}.
Decays of unstable particles
are described by \evtgen~\cite{Lange:2001uf}, in which final-state
radiation is generated using \photos~\cite{davidson2015photos}.
The interaction of the generated particles with the detector, and its response,
are implemented using the \geant
toolkit~\cite{Agostinelli:2002hh, *Allison:2006ve} as described in
Ref.~\cite{LHCb-PROC-2011-006}.

\section{Candidate selection}
\label{sec:Selection}

The $\decay{\Bs}{\jpsi\pip\pim}$ candidates are reconstructed by combining a $\jpsi$ candidate and two oppositely charged pions.
The $\jpsi$ candidate is
formed by two oppositely charged muons 
originating from a common vertex with good fit quality,
and each muon must have $\pt$ greater than $500\mevc$.
The invariant mass of the muon pair is required to be within \mbox{$[-48,\,+43]$}\mevcc of the known $\jpsi$ mass~\cite{PDG2024},
where the asymmetric mass window accounts for final-state radiation.
A $\pip\pim$ candidate is formed from two oppositely charged pions with $\pt$ greater than $250\mevc$,
originating from a common vertex with good fit quality that is significantly displaced from any PV.
Each pion is required to have $\chisqip$ greater than 4,
where $\chisqip$ is defined as the difference in the vertex-fit $\chisq$ of a given PV
when reconstructed with and without the track under consideration,
aiming to suppress tracks originating from a PV.
The scalar sum of the transverse momenta of these two pions is required to be greater than $900\mevc$.
All selected tracks must have
good fit quality, and pass loose particle identification (PID) criteria.
To suppress the candidates with one reconstructed track used more than once,
the opening angles between any two same-charge tracks are required to be greater than $0.5\mrad$. 
The $\jpsi$ and $\pip\pim$ candidates
are combined to form a $\Bs$ candidate.
The PV that fits best to the flight direction of the $\Bs$ candidate is taken as the associated PV.
The $\Bs$ candidate is required to have a decay vertex with good fit quality and to have
an invariant mass, $m(\jpsi\pip\pim)$, in the range of $\mbox{[5250,\,5580]}\mevcc$,
where $m(\jpsi\pip\pim)$ is calculated by requiring
the momentum of the $\Bs$ candidate to point back to the associated PV
and constraining the $\mup\mun$ invariant mass to the known $\jpsi$ mass~\cite{PDG2024}.
The decay time of the $\Bs$ candidate is required to be greater than $0.3\ps$, 
and its associated decay-time uncertainty to be less than $0.15\ps$. 
The candidates with one kaon or proton misidentified as a pion from $\decay{\Bds}{\jpsi\Kstarz(\Kp\pim)}$ and $\decay{\Lb}{\jpsi{p}\pim}$ decays, and from $\decay{\Bu}{\jpsi\Kp}$ decays 
combined with a random $\pim$ meson, are removed using combined PID and mass requirements,
whereas the candidates with two pions misidentified from $\decay{\Lb}{\jpsi{p}\Km}$ decay are suppressed.
The $\decay{\Bs}{\jpsi\pipm\pipm}$ sample with two pions of the same charge is selected with the same criteria and used to study the mass shape of the combinatorial background.

The $\decay{\Bs}{\jpsi\pip\pim}$ candidates in the mass range $[5250,\,5500]\mevcc$ are retained for further analysis.
A boosted decision tree (BDT) classifier~\cite{Breiman,AdaBoost}, 
implemented in the TMVA toolkit~\cite{Hocker:2007ht,*TMVA4}, 
is employed to further suppress the combinatorial background. 
The BDT classifier is trained using the simulated $\decay{\Bs}{\jpsi\pip\pim}$ decays as signal proxy and candidates in the $m(\jpsi\pip\pim)$ sideband region of $\mbox{[5500,\,5580]}\mevcc$ as background proxy. 
The variables used in the training are
the smaller log-likelihood difference between the muon and pion hypotheses for the two muons, 
the PID probability for each pion,
the scalar sum of the $\pt$ of pions,
the $\pt$ and direction angle of the $\Bs$ candidate, 
the natural logarithms of the maximum and minimum $\chisqip$ of the pions, 
the $\chisqip$ of the $\Bs$ candidate and the $\chisq$ of the $\Bs$ vertex and kinematic fit of the full decay chain~\cite{Hulsbergen:2005pu}.
The direction angle is defined as the angle between the $\Bs$ momentum vector and the line connecting the associated PV to the $\Bs$ decay vertex.
The simulation is corrected to match the distributions of
the minimum and maximum $\chisqip$ of the pions,
$\ptot$ and $\pt$ of the $\Bs$ candidate, the 
$\chisqip$ and $\chisq$ of kinematic fit,
the track multiplicity and the PID probabilities of pions
in background-subtracted data before the training.
The optimal requirement on the BDT response is obtained by maximising 
the quantity $\varepsilon / \sqrt{N_{\rm tot}} > 0.06$. 
The BDT selection efficiency, $\varepsilon$, is estimated with the signal simulation,
and $N_{\rm tot}$ is the total number of $\Bs$ candidates within $\pm 30\mevcc$ of the known $\Bs$ mass~\cite{PDG2024}. 

The events with more than one candidate, amounting to a small fraction of 2.8\%, are removed from the data sample.
The residual physics backgrounds:
the misidentified $\decay{\Lb}{\jpsi{p}\Km}$ decays, 
and the partially reconstructed $\decay{\Bs}{\jpsi\eta}(\pip\pim\g)$ and $\decay{\Bs}{\jpsi\etapr}(\rhoz\g)$ decays, where the photon is not reconstructed,
are studied using the corresponding simulated samples applying the same selections as those used for the data.

\section{Signal estimation and background subtraction}
\label{Sec:Signal}

A simultaneous unbinned maximum-likelihood fit is performed to the $m(\jpsi\pi\pi)$ distributions in the
$\decay{\Bs}{\jpsi\pip\pim}$ and $\decay{\Bs}{\jpsi\pipm\pipm}$ data samples.
The $\Bs$ and $\Bd$ peaks are described by a sum of a Hypatia function~\cite{Santos:2013gra} and a Gaussian function,
where the tail parameters of the Hypatia function and the fraction and width ratio relating the two functions are determined from simulations. 
The mass difference between the $\Bs$ and $\Bd$ peaks is constrained 
to the known value~\cite{PDG2024}. 
The mass shape of combinatorial background is modelled by a fifth-order Chebyshev polynomial,
whose parameters are shared between the $\decay{\Bs}{\jpsi\pip\pim}$ and $\decay{\Bs}{\jpsi\pipm\pipm}$ data samples.

The mass shapes of physics backgrounds,
including $\decay{\Lb}{\jpsi{p}\Km}$, $\decay{\Bs}{\jpsi\eta}$ and $\decay{\Bs}{\jpsi\etapr}$ decays,
are determined from the corresponding simulated samples. 
The yield of $\decay{\Lb}{\jpsi{p}\Km}$ decays is constrained to the value
extrapolated from the yield estimated using a fit to the $m(\jpsi{p}{\Km})$ distribution in the $\Bs$ sideband regions~\cite{LHCb-PAPER-2020-033},
where $m(\jpsi{p}\Km)$ is calculated with $\pip\pim$ candidates replaced by the ${p}\Km$ and $\antiproton\Kp$ hypotheses,
and that closest to the known $\Lb$ mass~\cite{PDG2024} is retained.
The yield ratio between $\decay{\Bs}{\jpsi\eta}$ and $\decay{\Bs}{\jpsi\etapr}$ decays, $N_{\jpsi\eta}/N_{\jpsi\etapr}$, is estimated through 
\begin{equation}
   \frac{N_{\jpsi\eta}}{N_{\jpsi\etapr}} = 
   \frac{\BF(\decay{\Bs}{\jpsi\eta})}{\BF(\decay{\Bs}{\jpsi\etapr})} \times
   \frac{\BF(\decay{\eta}{\pip\pim\gamma})}{\BF(\decay{\etapr}{\pip\pim\gamma})} \times
   \frac{\varepsilon_{\jpsi\eta}}{\varepsilon_{\jpsi\etapr}}\;,
   \label{eq_etaetapr}
\end{equation}
where $\BF$ denotes the branching fraction for each channel~\cite{PDG2024},
and $\varepsilon_{\jpsi\eta}$ ($\varepsilon_{\jpsi\etapr}$) is the total selection efficiency evaluated from the $\decay{\Bs}{\jpsi\eta}$ ($\decay{\Bs}{\jpsi\etapr}$) simulation. 
The yield ratio of $\decay{\Bs}{\jpsi\etapr}$ decay relative to the signal $\decay{\Bs}{\jpsi\pip\pim}$ decay, $N_{\jpsi\etapr}/N_{\jpsi\pip\pim}$, is estimated similarly.
These yield ratios 
are constrained to the values determined from the corresponding simulations.

\begin{figure}[tb]
    \includegraphics[width=0.5\linewidth]{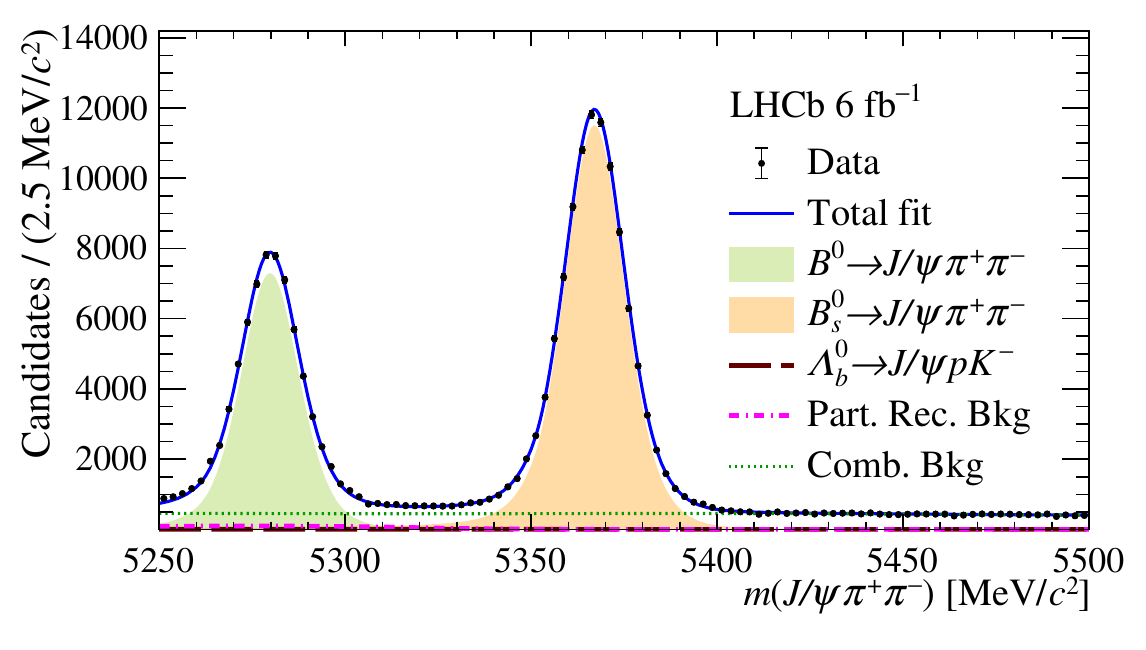}
    \includegraphics[width=0.5\linewidth]{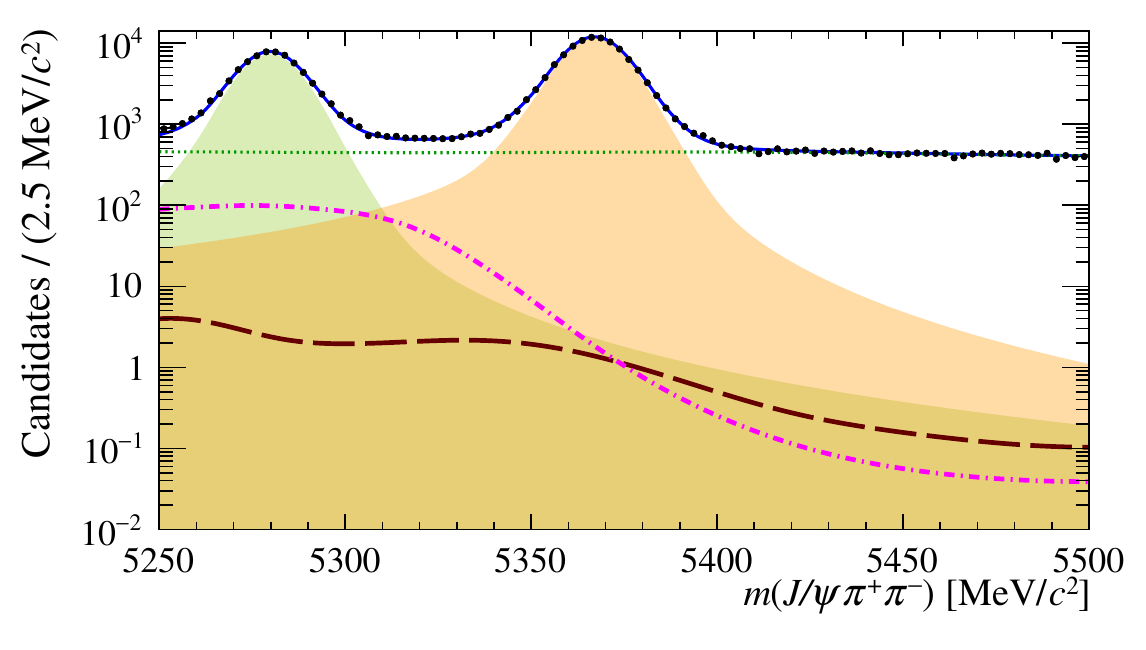}
  \caption{
  Distribution of the $J/\psi \pi^{+}\pi^{-}$ invariant mass in the full $\mbox{Run 2}\xspace$ data sample, shown with (left) linear and (right) logarithmic scales. 
  The total fit curve and contributions of $B^{0} \!\to J/\psi \pi^{+}\pi^{-}$, $B_{s}^{0} \!\to J/\psi \pi^{+}\pi^{-}$, $\Lambda_{b}^{0} \!\to J/\psi pK^{-}$,
  partially reconstructed background and combinatorial background are also presented.
  }
  \label{Fig2}
\end{figure}

The mass fit results, as shown in Fig.~\ref{Fig2},
are used to estimate the yield for each component within the signal region,
which is defined as $\pm 20\mevcc$ of the known $\Bs$ mass~\cite{PDG2024}. 
The $\Bs$ signal yields are estimated to be  
$32\,200 \pm 200$, $27\,950 \pm 190$ and $33\,800 \pm 200$
for the 2015--2016, 2017 and 2018 datasets, respectively. 
The contributions from $\decay{\Bd}{\jpsi\pip\pim}$, $\decay{\Bs}{\jpsi\eta}$, $\decay{\Bs}{\jpsi\etapr}$, and $\decay{\Lb}{\jpsi{p}\Km}$ decays
in the signal region
are found to be less than $0.1\%$, and are neglected.

%%%%%%%%%%%%%%%%%%%%%%%%%%%%%%%%%%%%%%%%%%%%%%%%
%%%%%%%%%%%%%%%%%%%%%%%%%%%%%%%%%%%%%%%%%%%%%%%%

To obtain a signal sample of $\decay{\Bs}{\jpsi\pip\pim}$ decays, 
the residual combinatorial background under the $\Bs$ peak 
is statistically subtracted by injecting background events with negative weights,
following the method in Ref.~\cite{LHCb-PAPER-2019-003}.
The injected events in the sideband range $\mbox{[5312,\,5332]}\mevcc$ are selected with a BDT requirement favouring the combinatorial background.

\section{Detector effects}
\label{sec:DetectorEffects}

The decay-time resolution, $\sigma_t$,
dominated by the uncertainty of the measured $\Bs$ flight distance, $\sigma_L$,
affects the precision of the \CP-violating parameters~\cite{LHCb-DP-2014-001}.
The per-candidate decay-time uncertainty, $\delta_t$,
derived from the momentum of the $\Bs$ candidate and the uncertainty on the flight distance, 
is imperfectly predicted.
Therefore, the estimated per-candidate decay-time uncertainty requires calibration.
This calibration is performed using a prompt $\jpsi\pip\pim$ sample, 
selected with identical criteria to the signal channel except for the requirements of $\chisqip$ of pions, and $\chisqip$ and decay time of $\Bs$ candidates. 
An unbinned maximum-likelihood fit is performed simultaneously to the decay-time distributions of the prompt $\jpsi\pip\pim$ data across ten subsamples, divided according to the value of decay-time uncertainty, which ensures sufficiently large subsamples.
The decay-time resolution in the fit is modelled using the sum of three Gaussian functions that share a common mean but have independent widths.
From the fit, the effective resolution is calculated using the widths and fractions of the Gaussian components.
The relation between the effective resolutions 
and the predicted decay-time uncertainties is then calibrated with a linear function.
Using the obtained calibration parameters and 
taking into account the distribution of decay-time uncertainty,
the average decay-time resolutions of the $\decay{\Bs}{\jpsi\pip\pim}$ decays 
are $40.23 \pm 0.05\fs$, $37.01\pm 0.05\fs$ and $36.93 \pm 0.05\fs$
for the 2015--2016, 2017 and 2018 datasets, respectively.
The uncertainties are statistical only.

The reconstruction and the selection induce a nonuniform efficiency as a function of the $\Bs$ decay time, decay angles, and the invariant $\pip\pim$ mass. 
The $\pip\pim$ mass and helicity angles share a common efficiency function, which is denoted as angular efficiency, 
and is assumed to be independent of the decay-time efficiency. 

The decay-time efficiency is determined using the control channel $\decay{\Bd}{\jpsi\Kstarz}$ in data and signal simulation. 
Before the efficiency is determined,
the $\decay{\Bs}{\jpsi\pip\pim}$ simulation is weighted
according to the amplitude model in 
Ref.~\cite{LHCb-PAPER-2019-003},
and then corrected to match the distributions of the $\Bs$ transverse momentum and pseudorapidity, and the track multiplicity in background-subtracted data.
The total weight, which is the product of the amplitude and correction weights, in the corrected signal simulation is used in the determination of the decay-time efficiency.
The $\decay{\Bd}{\jpsi\Kstarz}$ simulation is corrected to match the distributions of the $\Bd$ transverse momentum and pseudorapidity, the track multiplicity, the $\Kp\pim$ invariant mass, and the helicity angle $\cos\theta_{K\pi}$ in background-subtracted $\decay{\Bd}{\jpsi\Kstarz}$ data.
The decay-time efficiency of the signal candidates
is calculated as $\varepsilon^{\Bs}(t) = \varepsilon^{\Bd}_{\rm data}(t)
\times \varepsilon^{\Bs}_{\rm sim}(t)/\varepsilon^{\Bd}_{\rm sim}(t)$~\cite{LHCb-PAPER-2019-003},
where $\varepsilon_{\rm data}^{\Bd}(t)$ is the efficiency of $\decay{\Bd}{\jpsi\Kstarz}$ decay determined from the data, and
$\varepsilon_{\rm sim}^{\Bs}(t)$ ($\varepsilon_{\rm sim}^{\Bd}(t)$) is the efficiency of the $\decay{\Bs}{\jpsi\pip\pim}$ ($\decay{\Bd}{\jpsi\Kstarz}$) decay determined from the simulation.
The ratio $\varepsilon_{\rm sim}^{\Bs}(t)/\varepsilon_{\rm sim}^{\Bd}(t)$ accounts for the kinematic differences between the two modes.
All these efficiencies are modelled using spline functions and are determined via a simultaneous fit to the decay-time distributions in the 
control data sample, and signal and control simulated samples. 
The decay-time distribution for each sample is described by an exponential function convolved with a Gaussian resolution function.
For $\decay{\Bd}{\jpsi\Kstarz}$ samples,
the lifetime of $\Bd$ is fixed to its known value~\cite{PDG2024},
and the parameters of the resolution function are fixed to the values obtained from the resolution study.
The decay-time efficiencies are determined independently for both trigger categories.

The angular efficiency is determined from the $\decay{\Bs}{\jpsi\pip\pim}$ simulation,
which is corrected to match the distributions of the minimum and maximum $\chisqip$ of the pions,
the $\Bs$ momentum, transverse momentum, $\chisqip$ and the $\chisq$ of the kinematic fit~\cite{Hulsbergen:2005pu},
the track multiplicity and the PID probabilities of pions in background-subtracted data.
Since the simulation is generated with an equal mixture of the $\Bs$ mass eigenstates 
and the data is dominated by the heavy mass eigenstate,
an extra weight of $e^{-t/\tauH}/\cbracket{(e^{-t/\tauH} + e^{-t/\tauL})/2}$ is applied 
to correct this difference between data and simulation,
where $\tauH$ ($\tauL$) is the lifetime of heavy (light) $\Bs$ eigenstate.
The angular efficiency is described by a combination of 
the Legendre polynomials and the spherical harmonic basis functions as~\cite{LHCb-PAPER-2017-008, LHCb-PAPER-2019-003}
\begin{equation}
\varepsilon(\mpipi,\,\boldsymbol{\Omega})=
\sum_{a,\,b,\,c,\,d}\varepsilon^{abcd}P_{a}(\cospi)Y_{b}^{c}(\cosmu,\chi)
P_{d}\mbracket{2 r_{\pi\pi}-1}\;,
\end{equation}
where $\varepsilon^{abcd}$ are the angular coefficients, $P_a$ and $P_d$ are Legendre polynomials, $Y_{b}^{c}$ are spherical harmonic basis functions, 
the quantity $r_{\pi\pi}$ is defined as
$(\mpipi-2m_{\pi})/(m_{\Bs}-m_{\jpsi}-2m_{\pi})$.
The relation $\varepsilon^{abcd}=(-1)^c\varepsilon^{ab(-c)d}$ is required to ensure the resulting angular efficiency is real.
Following the procedure detailed in Ref.~\cite{LHCb-PAPER-2017-008},
the coefficients of the angular efficiencies are obtained in a finite range with $a\leq10$, $b\leq8$, $\abs{c}\leq2$, $d\leq8$,
separately for the two trigger categories,
and retained if the statistical significance is greater than three standard deviations from zero.

\section{Flavour tagging}
\label{Sec:FlavTag}

The determination of the initial flavour for a $\BsorBsbar$ meson at production
is necessary for measuring 
the $\CP$-violating phase.
This is achieved through
the opposite-side (OS) and the same-side (SS) taggers~\cite{Fazzini:2018dyq, LHCb-PAPER-2011-027, LHCb-PAPER-2015-027, LHCb-PAPER-2015-056}.
The OS tagger infers the flavour of a signal $\bquark$-hadron
with the information from the decays of the opposite $\bquark$-hadron in the event, 
while the SS tagger does this with the information from the fragmentation products of the signal $\bquark$-hadron.
Each algorithm assigns a per-candidate tagging decision, $q$,
and a mistag probability, $\eta$, based on the output of its neural network.
Here $q$ takes values of $+1$, $-1$ and~$0$,
corresponding to the classification as $\Bs$, $\Bsb$ and ${\rm untagged}$, respectively.
Due to the specific training and optimisation of the simulation samples used for each algorithm,
the per-candidate mistag probability is imperfectly estimated.
Following the procedure in Ref.~\cite{LHCb-PAPER-2019-003}, the corrected mistag probability, $\omega$ ($\overline{\omega}$), as a function of the mistag probability for $\Bs$ ($\Bsb$) assigned by the algorithm is calibrated using a linear model for $\Bs$ ($+$ sign) and $\Bsb$ ($-$ sign), 
\begin{equation}
    {\ensuremath{\kern \thebaroffset\optbar{\kern -\thebaroffset \omega}}\xspace}(\eta)
     = \mbracket{p_0  \pm \frac{\Delta p_0}{2}} +
     \mbracket{p_1 \pm \frac{\Delta p_1}{2}}
     \mbracket{\eta - \langle\eta\rangle}\;,
\label{eqn_tagging_calib}
\end{equation}
where $p_0$ and $p_1$ are the calibration parameters,
and $\Delta p_0$ and $\Delta p_1$ account for the tagging asymmetry between $\Bs$ and $\Bsb$ mesons.

The calibrations of the OS tagger and SS tagger are performed using control samples of the charged $\decay{\Bu}{\jpsi\Kp}$ decays and the flavour-specific $\decay{\Bs}{\Dsm\pip}$ decays, respectively. 
These samples are corrected to match the distributions 
of the transverse momentum and pseudorapidity of $\B$ ($\Bu$ and $\Bs$) candidate, the track multiplicity, and the number of PVs in the signal data sample. 
The OS calibration is performed using the number of candidates with correct and wrong decisions in the $\decay{\Bu}{\jpsi\Kp}$ data,
while the SS calibration is determined from a fit to the decay-time distribution of $\decay{\Bs}{\Dsm\pip}$ data.

The effective tagging power, defined as the product of the tagging efficiency, $\varepsilon_{\rm tag}$, and the average squared dilution, $\langle D^2\rangle = \langle(1-2\omega)^2\rangle$,
quantifies the effective tagged fraction of the signal events.
The resulting average tagging powers of the $\decay{\Bs}{\jpsi\pip\pim}$ data samples are summarised in Table~\ref{tab_TagPow}. 

%%%%%%%%%%%%%%%%%%%%%%%%%%%%%%%%%
\begin{table}[tb]
    \begin{center}
	\caption{
    Tagging powers ($\varepsilon_{\rm tag}\langle D^2\rangle$, in\,\%) of $\decay{\Bs}{\jpsi\pip\pim}$ data for different years.
    OS-only labels candidates tagged only by the opposite-side tagger,
    SS-only labels candidates tagged only by the same-side tagger,
    OS\&SS labels candidates tagged by both taggers,
    and total is the sum of all categories.
    The uncertainties are statistical only.
    }
	\label{tab_TagPow}
	\begin{tabular}{c r@{\:$\pm$\:}l r@{\:$\pm$\:}l r@{\:$\pm$\:}l }
	   \hline
	   Category    
	   & \multicolumn{2}{c}{2015--2016}
	   & \multicolumn{2}{c}{2017}
	   & \multicolumn{2}{c}{2018} \\
       \hline
	   OS-only & 0.88 & 0.04 & 0.85 & 0.05 & 0.91 & 0.04 \\
       SS-only & 1.02 & 0.16 & 1.02 & 0.14 & 1.21 & 0.18 \\
       OS\&SS  & 2.50 & 0.11 & 2.59 & 0.13 & 2.63 & 0.12 \\
       \hline
       Total   & 4.40 & 0.20 & 4.47 & 0.20 & 4.76 & 0.22 \\
       \hline
	\end{tabular}
   \end{center}
\end{table}
%%%%%%%%%%%%%%%%%%%%%%%%%%%%%%%%%

\section{Resonant structure of the \texorpdfstring{\boldmath{$\pip\pim$}}{pi+pi-} spectrum}
\label{Sec:Resonances}

The amplitude of the resonance $R$ in the $\mpipi$ spectrum is expressed as~\cite{Zhang:2012zk,LHCb-PAPER-2013-069,LHCb-PAPER-2019-003}
\begin{equation}
    \label{eqn_amplitude}
    {\cal A}_R(\mpipi) = \sqrt{2J_R+1}\sqrt{P_RP_{\Bs}}
    F_R^{(L_R)}F_{\Bs}^{(L_{\Bs})}{\cal M}_R(\mpipi)
    \mbracket{\frac{P_R}{m_R}}^{L_R}
    \mbracket{\frac{P_{\Bs}}{m_{\Bs}}}^{L_{\Bs}}\;,
\end{equation}
where ${\cal M}_R(\mpipi)$ is the Flatt\'e function or the relativistic Breit--Wigner (RBW) function.
In this equation, $J_R$ and $m_R$ are the spin and pole mass of each resonance,
$P_{R}$ ($P_{\Bs}$) is the momentum of the $\pip$ ($\jpsi$) meson in the $\pip\pim$ ($\Bs$) 
rest frame, 
$L_R$ is the orbital angular momentum between $\pip$ and $\pim$ mesons,
$L_{\Bs}$ is the orbital angular momentum between the $\jpsi$ meson and the $\pip\pim$ system,
and $F_R^{(L_R)}$ and $F_{\Bs}^{(L_{\Bs})}$ are the Blatt--Weisskopf barrier factors for resonances and $\Bs$ mesons~\cite{Blatt:1952ije}, respectively.

The determination of the \CP-violating phase, $\phis$,
is performed on the entire $\pip\pim$ mass spectrum.
In the previous analyses, five resonances were observed in the $\pip\pim$ spectrum~\cite{LHCb-PAPER-2012-005,LHCb-PAPER-2013-069,LHCb-PAPER-2019-003}, namely
$f_0(980)$, $f_2(1270)$, $f_0(1500)$, $f_2^\prime(1525)$ and $f_0(1770)$,
along with a nonresonant (NR) component.
In this analysis,
an extra resonant contribution from $f_2(1565)$ is included to describe a small peak around~$1.6\gevcc$.
The $f_0(980)$ resonance is described using a Flatt\'e function~\cite{Flatte:1976xv}, 
while the other resonances are modelled by RBW functions~\cite{LHCb-PAPER-2018-005}.
The values of pole masses and decay widths of the RBW functions are taken from Particle Data Group (PDG)~\cite{PDG2024},
\lhcb~\cite{LHCb-PAPER-2017-008}, \obelix~\cite{OBELIX:1998nvy}
and \bes~\cite{BES:2004twe},
as summarised in Table~\ref{tab_infor_resonance}. 
The nonresonant component is modelled flat across the phase space.

%%%%%%%%%%%%%%%%%%%%%%%%%%%%%%%%%
\begin{table}[tb]
  \begin{center}
 \caption{
    The pole masses and decay widths of the $\pip\pim$ resonances. 
      The values are taken from PDG~\cite{PDG2024}, \lhcb~\cite{LHCb-PAPER-2017-008}, \obelix~\cite{OBELIX:1998nvy} and \bes~\cite{BES:2004twe}.
      }
 \label{tab_infor_resonance}
 \begin{tabular}{ccccc}
    \hline
    Resonance & Model & Mass [\!\mevcc] 
    & Width [\!\mevcc] & Reference \\
    \hline
      $f_{0}(980)$  & Flatt\'e & -- & -- & \\
    $f_{2}(1270)$ & RBW    & $1275.4 \pm 0.8$  & $186.6^{\:+\:2.8}_{\:-\:2.2}$ & PDG~\cite{PDG2024}  \\
      $f_{0}(1500)$ & RBW    & -- & -- &  \\
      $f_{2}^{\prime}(1525)$ & RBW & $1522.2 \pm 1.7$  & $78.0 \pm 4.8$ & \lhcb~\cite{LHCb-PAPER-2017-008}  \\
      $f_{2}(1565)$ & RBW    & $1575 \pm 18$   & $119 \pm 24$  & \obelix~\cite{OBELIX:1998nvy} \\
      $f_{0}(1770)$ & RBW    & $1790^{\:+\:40}_{\:-\:30}$   & $270^{\:+\:60}_{\:-\:30}$ & \bes~\cite{BES:2004twe}  \\
    \hline
 \end{tabular}
  \end{center}
\end{table}
%%%%%%%%%%%%%%%%%%%%%%%%%%%%%%%%%%

\section{Maximum-likelihood fit}
\label{sec:Results}

A simultaneous unbinned maximum-likelihood fit is performed to six datasets,
divided into the three periods of data-taking 
and the two trigger categories,
to obtain the \CP-violation parameters in the $\decay{\Bs}{\jpsi\pip\pim}$ decays. 
All candidates tagged by the OS or SS tagger are retained for the fit.
For the candidates tagged by both taggers, 
a combined tag decision, $\mathfrak{q}$, and 
a combined mistag probability, $\mathfrak{y}$,
are established using the method in Ref.~\cite{LHCb-PAPER-2011-027}.

The fit minimises the negative log-likelihood function, expressed as~\cite{LHCb-PAPER-2019-003,LHCb-PAPER-2017-008}
\begin{equation}
    \label{eqn_log_likelihood}
    -2\ln{\cal L} = -2\alpha\sum_i w_i\ln\pdfp(t,\,\mpipi,\,\boldsymbol{\Omega},\,\mathfrak{q}|\mathfrak{y},\,\delta_t;\,\boldsymbol{a})\;,
\end{equation}
where $i$ runs over all candidates,
$w_i$ is the per-candidate weight providing the background subtraction, 
the factor ${\alpha = \sum_i w_i/\sum_i w_i^2}$ accounts for the statistical effect of the background subtraction~\cite{Xie:2009rka},
and $\pdfp(t,\,\mpipi,\,\boldsymbol{\Omega},\,\mathfrak{q}|\mathfrak{y},\,\delta_t;\,\boldsymbol{a})$ is the signal probability density function (PDF).
This five-dimensional PDF describes the distributions for $\Bs$ and $\Bsb$ decays
of the reconstructed decay time, $t$, 
the $\pip\pim$ invariant mass, $\mpipi$,
and the helicity angles, $\boldsymbol{\Omega}=(\cospi,\,\cosmu,\,\chi)$.
It is given as
\begin{equation}
\pdfp(t,\,\mpipi,\,\boldsymbol{\Omega},\,\mathfrak{q}|\mathfrak{y},\,\delta_t;\,\boldsymbol{a}) = \frac{1}{\cal N}\cbracket{S(\tprime,\,\mpipi,\,\boldsymbol{\Omega},\,\mathfrak{q}|\mathfrak{y};\,\boldsymbol{a})\otimes G(t-t^\prime|\delta_t)}\varepsilon^{\Bs}(t)\,\varepsilon(\mpipi,\,\boldsymbol{\Omega})\;,
\end{equation}
where $\boldsymbol{a}$ is a set of physics parameters to be measured,
including the \CP-violating phase, $\phis$,
the direct \CP-violation parameter, $\abs{\lambda}$,
the decay width of the heavy mass eigenstate, $\GH$,
and the moduli and phases of the amplitudes of the $\pip\pim$ resonant and nonresonant components discussed in Sec.~\ref{Sec:Resonances}.
The factor ${\cal N}$ accounts for the overall normalisation
and $\tprime$ is the true decay time.
$S(\tprime,\,\mpipi,\,\boldsymbol{\Omega},\,\mathfrak{q}|\mathfrak{y};\,\boldsymbol{a})$ is the decay-time distribution including the flavour tagging effects,
$G(t-\tprime|\delta_t)$ is a Gaussian function modelling the decay-time resolution,
and $\varepsilon^{\Bs}(t)$ and $\varepsilon(\mpipi,\,\boldsymbol{\Omega})$ are the decay-time and angular efficiencies,
obtained in Sec.~\ref{sec:DetectorEffects}.
The decay-time distribution is 
\begin{equation}
    \begin{aligned}
    S &(\tprime,\,\mpipi,\,\boldsymbol{\Omega},\,\mathfrak{q}|\mathfrak{y};\,\boldsymbol{a})  = 
\cbracket{ \varepsilon_{\rm tag}\mbracket{\frac{\abs{\mathfrak{q}}}{2}+\frac{\mathfrak{q}}{2}(1-2\omega(\mathfrak{y}))}
    +(1-\varepsilon_{\rm tag})(1-\abs{\mathfrak{q}})} \Gamma(\tprime,\,\mpipi,\,\boldsymbol{\Omega};\boldsymbol{a})\\
& +\cbracket{ \varepsilon_{\rm tag}\mbracket{\frac{\abs{\mathfrak{q}}}{2}-\frac{\mathfrak{q}}{2}(1-2\overline{\omega}(\mathfrak{y}))}
    +(1-\varepsilon_{\rm tag})(1-\abs{\mathfrak{q}})}\frac{1+A_{\rm P}}{1-A_{\rm P}} \overline{\Gamma}(\tprime,\,\mpipi,\,\boldsymbol{\Omega};\boldsymbol{a})\;,
\end{aligned}
\end{equation}
where $A_{\rm P}$ is the production asymmetry of $\Bs$ mesons,
and $\GorGb(\tprime,\,\mpipi,\,\boldsymbol{\Omega};\boldsymbol{a})$ are the time-dependent decay rates for $\decay{\BsorBsbar}{\jpsi\pip\pim}$ decays defined in Eq.~\ref{eq:time-dependent-decay-rate}.

The pole masses and decay widths of the $f_{2}(1270)$ and $f_{2}^{\prime}(1525)$ resonances are fixed to the values in Table~\ref{tab_infor_resonance}, 
and those of the $f_2(1565)$ and $f_0(1770)$ resonances are constrained. 
The parameters of the $f_0(980)$ and $f_0(1500)$ resonances are varied freely.
The amplitude modulus and phase of the $f_0(980)$ resonance are fixed to 1 and 0, respectively, 
to serve as a reference, while those of other components are varied freely.
The calibration parameters of the decay-time resolution,
the coefficients of the angular and decay-time efficiencies
are fixed to the values obtained in Sec.~\ref{sec:DetectorEffects}.
The calibration parameters of the flavour taggers are constrained to the values 
obtained in Sec.~\ref{Sec:FlavTag},
taking into account their correlations. 
The mass difference, $\dms$, and 
the decay width of the light mass eigenstate, $\GL$,
are fixed to the known value~\cite{PDG2024}.
The production asymmetry, $A_{\rm P}$, is fixed to zero, following Ref.~\cite{LHCb-PAPER-2019-003}.

%%%%%%%%%%%%%%%%%%%%%%%%%%%%%%%%%%%%%%%%%%%
\begin{table}[tb]
  \begin{center}
  \caption{
    Physics parameters of interest, including the \CP-violating phase, $\phis$, the direct \CP-violation parameter, $\abs{\lambda}$, and 
    the decay width of the heavy mass eigenstate, $\GH$,
    along with the correlations.
    The first uncertainty is statistical and the second systematic,
    as discussed in Sec.~\ref{sec:Systematics}.
 }
 \label{tab_bscpv}
 \begin{tabular}{l r@{\:$\pm$\:}c@{\:$\pm$\:}l ccc}
    \hline
    \multirow{2}{*}{Parameter}                 &  \multicolumn{3}{c}{\multirow{2}{*}{Value with uncertainties}}  & \multicolumn{3}{c}{Correlation}   \\
    {} & \multicolumn{3}{c}{} & $\phis$ & $\abs{\lambda}$ & $\GH$  \\
    \hline
    $\phis$ [$\rad$]    &$-$0.077 & 0.034 & 0.007 & 1\phantom{.000} & 0.073 & 0.035 \\
    $\abs{\lambda}$     &   0.993 & 0.026 & 0.007 & 0.073 & 1\phantom{.000} & 0.023 \\
    $\GH$ [$\invps$]    &   0.610 & 0.002 & 0.004 & 0.035 & 0.023 & 1\phantom{.000} \\
 \hline
 \end{tabular}
  \end{center}
\end{table}
%%%%%%%%%%%%%%%%%%%%%%%%%%%%%%%%%%%%%%%%%%%%
\begin{figure}[tb]
  \begin{center}
     \includegraphics[width=0.90\linewidth]{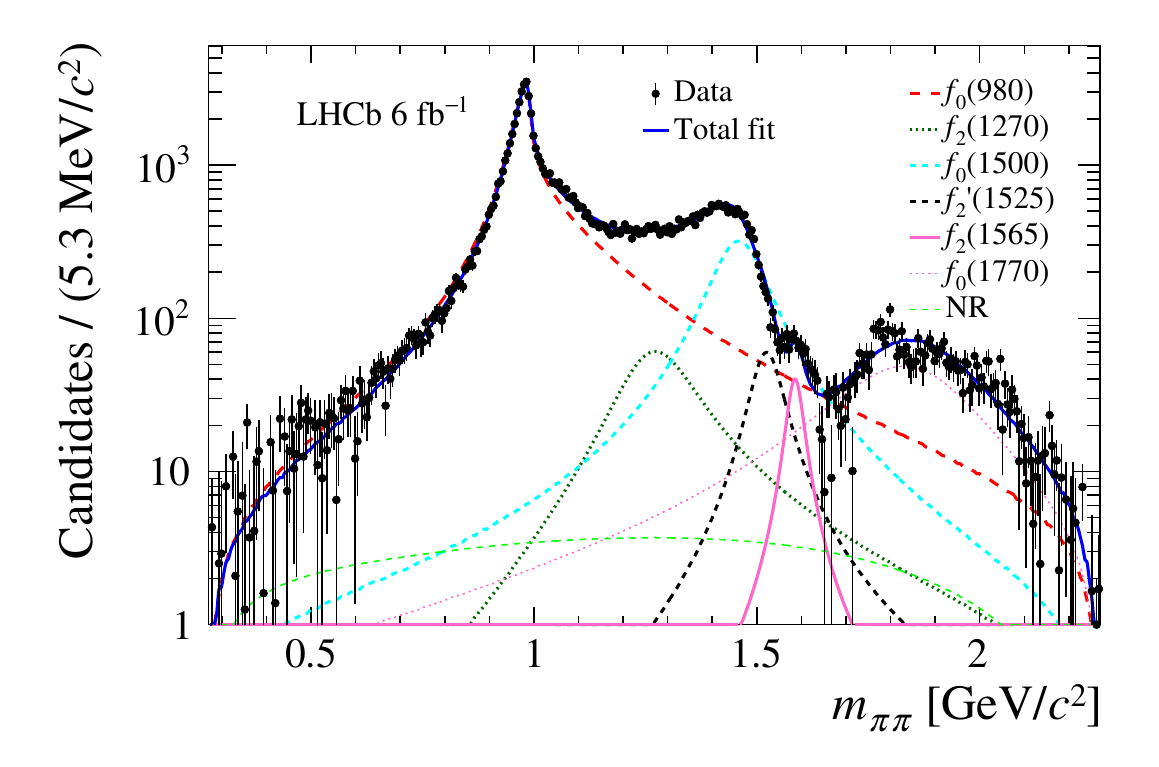}
     \vspace*{-0.5cm}
  \end{center}
  \caption{%\small
     Distribution of the $\pi^{+}\pi^{-}$ mass spectrum in the full $\mbox{Run 2}\xspace$ data sample. The total fit curve and the resonant and nonresonant contributions are also shown.
     }
  \label{Fig3}
\end{figure}
%%%%%%%%%%%%%%%%%%%%%%%%%%%%%%%%%%%%%%%%%%%%
\begin{figure}[tb]
  \begin{center}
     \includegraphics[width=0.45\linewidth]{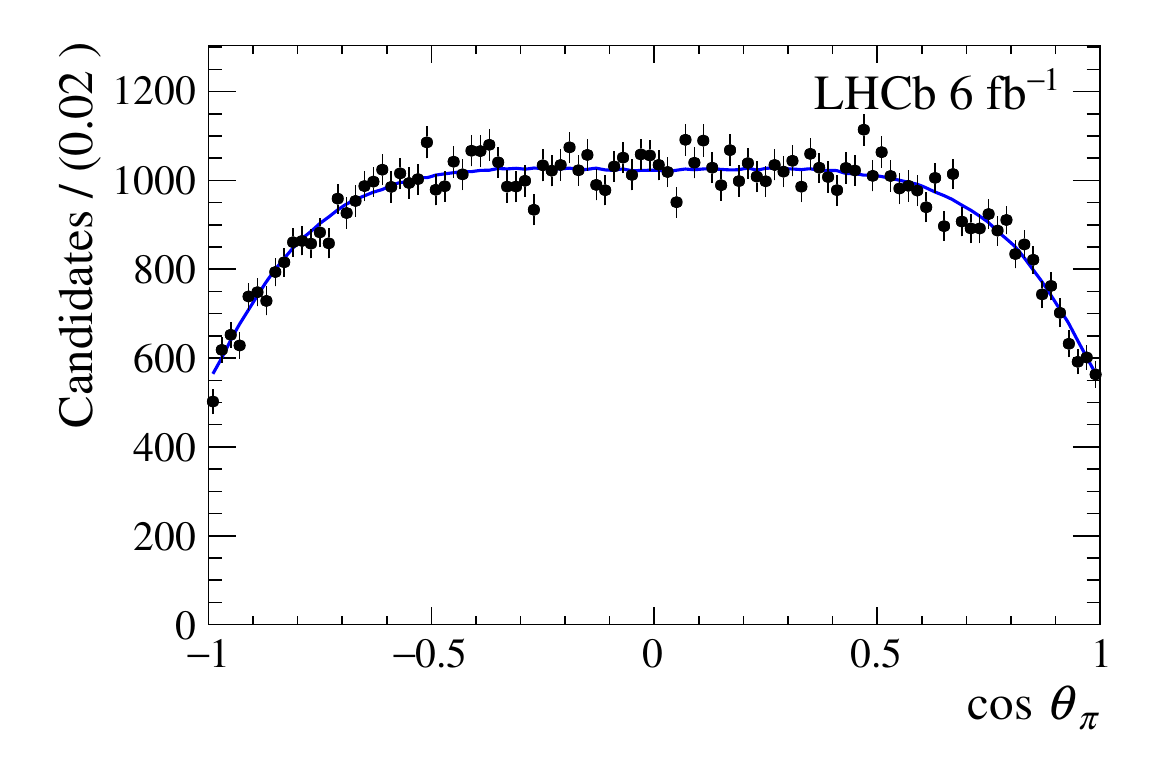}
     \includegraphics[width=0.45\linewidth]{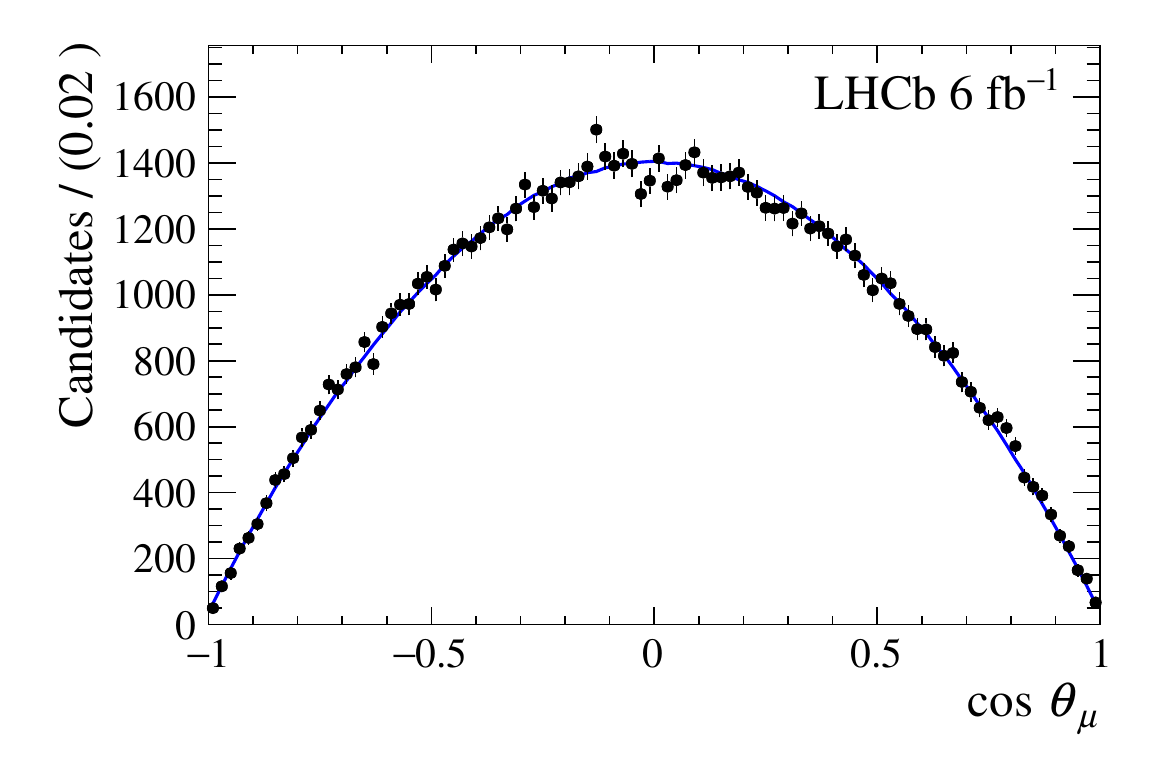}
     \includegraphics[width=0.45\linewidth]{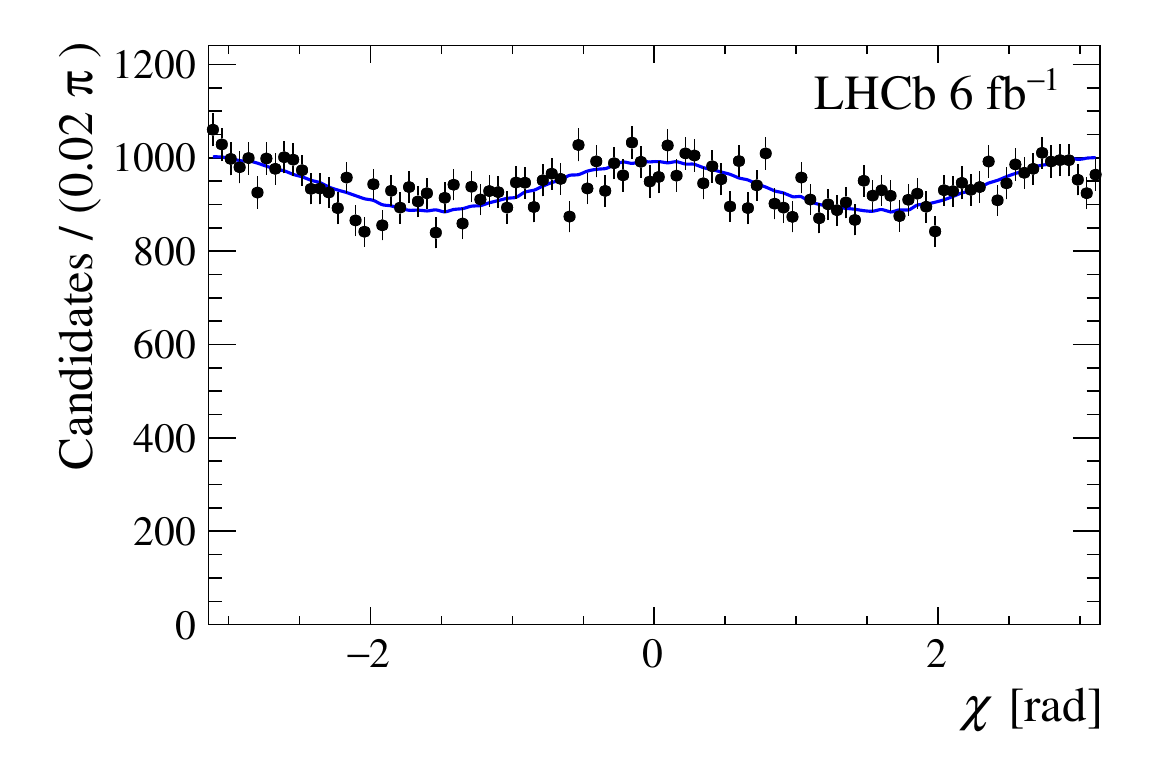}
     \includegraphics[width=0.45\linewidth]{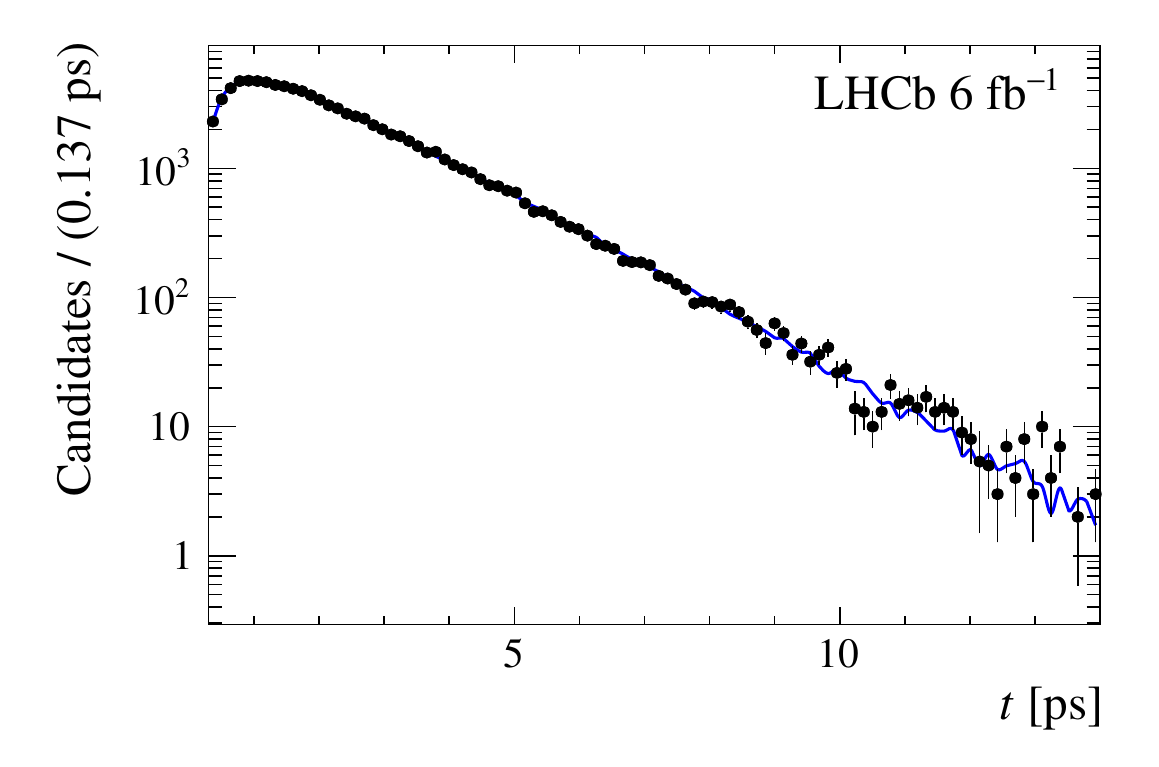}
     \vspace*{-0.5cm}
  \end{center}
  \caption{
     Distributions of 
     (top left) $\cos\theta_\pi$, (top right) $\cos\theta_\mu$, (bottom left) $\chi$ and (bottom right)~$t$.
     The points with error bars represent the data,
     and the solid blue line the fit curve.
    }
  \label{Fig4}
\end{figure}
%%%%%%%%%%%%%%%%%%%%%%%%%%%%%%%%%%%%%%%%%%%%%

The physics parameters of interest determined from the fit are summarised in Table~\ref{tab_bscpv}, along with their correlations.
These results are consistent with those reported in previous analyses~\cite{LHCb-PAPER-2014-019,LHCb-PAPER-2019-003}. 
As a cross-check, the fit is repeated on the subsets of data
divided by data-taking years and magnet polarities, and the obtained results are consistent with the baseline results within 1.5 statistical deviations in all cases.
The background-subtracted distributions of $\mpipi$, decay angles and decay time, along with fit projections, are shown in Figs.~\ref{Fig3} and~\ref{Fig4}, respectively. 
The fit quality is assessed using the $\chisq$ test,
yielding the $\chisqndf$ values of 490/378, 105/99, 74/99, 104/99, 123/99 for
$\mpipi$, $\cospi$, $\cosmu$, $\chi$ and $t$, respectively.
A goodness-of-fit test using the $k$-dimensional tree method~\cite{Williams:2010vh} is applied to the background-subtracted data sample,
and yields $\chisqndf = 2357/2056$.

The contribution of each resonant and nonresonant component
is quantified by the fit fraction, 
${\cal F}_i = \int\abs{{\cal A}_i}^2{\rm d}\mpipi{\rm d}\boldsymbol{\Omega}/\int\abs{\cal A_{\rm tot}}^2{\rm d}\mpipi{\rm d}\boldsymbol{\Omega}$, 
where the numerator is the integral of the squared amplitude of component $i$ over the phase space and the denominator is that of the total squared amplitude. 
The obtained fit fractions for each component, along with the transversity fractions,  
and the corresponding amplitude parameters 
are summarised in Tables~\ref{tab_pipi_frac} and \ref{tab_pipi_amp}, respectively.
The interference among different resonances leads to the sum of the fit fraction being less than 100\%.

%%%%%%%%%%%%%%%%%%%%%%%%%%%%%%%%%%%%%%%%%%%%
\begin{table}[tb]
  \begin{center}
 \caption{
    Fit fractions and transversity fractions for each $\pip\pim$ component.
    The uncertainties are statistical only.  
 }
 \label{tab_pipi_frac}
 \begin{tabular}{l r@{\:$\pm$\:}l r@{\:$\pm$\:}l r@{\:$\pm$\:}l r@{\:$\pm$\:}l }
    \hline
    \multirow{2}{*}{Component}
    & \multicolumn{2}{c}{\multirow{2}{*}{Fit fraction (\%)}}
    & \multicolumn{6}{c}{Transversity fractions (\%)} \\ 
    {}
    & \multicolumn{2}{c}{} & \multicolumn{2}{c}{$0$}
    & \multicolumn{2}{c}{$\parallel$} & \multicolumn{2}{c}{$\perp$} \\
    \hline
    $f_{0}(980)$ & 72.5 & 1.9 & \multicolumn{2}{c}{100} & \multicolumn{2}{c}{0} & \multicolumn{2}{c}{0} \\
    $f_{0}(1500)$ & 12.6 & 0.7 & \multicolumn{2}{c}{100} & \multicolumn{2}{c}{0} & \multicolumn{2}{c}{0} \\
    $f_{2}(1270)$ & 3.2 & 0.2 & 19.8 & 1.9 & 26.6 & 4.3 & 53.6 & 4.5 \\
    $f_{2}^{\prime}(1525)$ & 1.4 & 0.2 & 51.5 & 7.1 & 13.2 & 4.2 & 35.3 & 8.6 \\
    $f_{2}(1565)$ & 0.5 & 0.1 & 12.0 & 3.4 & 20.0 & 8.6 & 68.1 & 9.1 \\
    $f_{0}(1770)$ & 4.4 & 0.6 & \multicolumn{2}{c}{100} & \multicolumn{2}{c}{0} & \multicolumn{2}{c}{0} \\
    NR & 1.0 & 0.3 & \multicolumn{2}{c}{100} & \multicolumn{2}{c}{0} & \multicolumn{2}{c}{0} \\
    \hline
  \end{tabular}
  \end{center}
\end{table}
%%%%%%%%%%%%%%%%%%%%%%%%%%%%%%%%
\begin{table}[tb]
    \begin{center}
	\caption{
	   Relative amplitude moduli, $\abs{A}$, and phase differences, $\delta$, of $\pip\pim$ resonant and nonresonant contributions.
       The uncertainties are statistical only.
       The amplitude moduli are ratios relative to the reference state,
       and the phases are differences relative to the same reference.
	}
    \label{tab_pipi_amp}
	\begin{tabular}{ c c r@{\:$\pm$\:}l r@{\:$\pm$\:}l }
	   \hline
	   Component & Reference & \multicolumn{2}{c}{\abs{A}} & \multicolumn{2}{c}{$\delta$ (${}^\circ$)} \\ 
	   \hline
	        $f_{2}(1270)_{0}$ & $f_{0}(980)_{0}$ & 0.95 & 0.05 & 111 & 5 \\ 
        $f_{2}(1270)_{\perp}$ & $f_{0}(980)_{0}$ & 1.57 & 0.11 & 298 & 21 \\ 
    $f_{2}(1270)_{\parallel}$ & $f_{0}(980)_{0}$ & 1.11 & 0.09 & 290 & 6 \\ 
            $f_{0}(1500)_{0}$ & $f_{0}(980)_{0}$ & 0.45 & 0.02 & 161 & 7 \\ 
   $f_{2}^{\prime}(1525)_{0}$ & $f_{0}(980)_{0}$ & 0.80 & 0.05 & 270 & 8 \\ 
$f_{2}^{\prime}(1525)_{\perp}$ & $f_{2}(1270)_{\perp}$ & 0.66 & 0.13 & 208 & 9 \\ 
 $f_{2}^{\prime}(1525)_{\parallel}$ & $f_{0}(980)_{0}$ & 0.40 & 0.07 & 129 & 9 \\ 
            $f_{2}(1565)_{0}$ & $f_{0}(980)_{0}$ & 0.18 & 0.03 & 237 & 12 \\ 
   $f_{2}(1565)_{\perp}$ & $f_{2}(1270)_{\perp}$ & 0.43 & 0.07 & 135 & 13 \\ 
    $f_{2}(1565)_{\parallel}$ & $f_{0}(980)_{0}$ & 0.23 & 0.05 & 159 & 20 \\ 
            $f_{0}(1770)_{0}$ & $f_{0}(980)_{0}$ & 0.77 & 0.09 & 129 & 17 \\ 
            $\mathrm{NR}_{0}$ & $f_{0}(980)_{0}$ & 0.24 & 0.04 & 350 & 16 \\
	   \hline
	\end{tabular}
   \end{center}
\end{table}
%%%%%%%%%%%%%%%%%%%%%%%%%%%%%%%%%%%%%%%%%%%

\section{Systematic uncertainties}
\label{sec:Systematics}

The sources of systematic uncertainties are summarised in Table~\ref{tab_syst}. 
They are evaluated either by repeating the maximum-likelihood fit
with a modified model or parameter and observing the changes in the physics parameters,
or by performing pseudoexperiments to estimate the uncertainties due to limited sample sizes.

%%%%%%%%%%%%%%%%%%%%%%%%%%%%%%%%
\begin{table}[tb]
\begin{center}
  \caption{%\small
    Systematic uncertainties from different sources. 
    The total systematic uncertainties and statistical uncertainties are listed.
    The values with ``--" have a value of less than 0.05.
  }
  \label{tab_syst}
  \begin{tabular}{l S[table-format=2.1] S[table-format=2.1] S[table-format=2.1]}
  \hline
  Source 
  & {\tabincell{c}{ $\phis$ \\ $[\mrad\,]$ }}     
      & {\tabincell{c}{ $\abs{\lambda}$ \\ $[10^{-3}\,]$ }} 
  & {\tabincell{c}{ $\Gamma_{\rm H}$ \\ $[\invfs\,]$}} \\
  \hline  
    $m(\jpsi\pip\pim)$ fit model         	    & \multicolumn{1}{c}{--}  & 0.1 & 0.1 \\
    Resonance model                             & 3.4 & 0.9 & 0.1 \\
    Width of light eigenstate $\GL$             & 0.6 & 0.2 & 0.1 \\
    Mass difference $\dms$                      & 1.0 & 0.8 & \multicolumn{1}{c}{--}  \\
    Resonance mass and width               	    & 0.2 & 0.4 & \multicolumn{1}{c}{--}  \\
    Production asymmetry $A_{\rm P}$            & 3.2 & 0.5 & \multicolumn{1}{c}{--}  \\
    Decay-time resolution scale                 & 3.8 & 1.9 & \multicolumn{1}{c}{--}  \\
    Decay-time bias correction                  & 3.8 & 6.8 & \multicolumn{1}{c}{--}  \\
    Decay-time resolution calculation       	& 0.8 & 0.4 & \multicolumn{1}{c}{--}  \\
    Decay-time efficiency                  	    & 0.5 & 0.1 & 3.6 \\
    Decay-time efficiency spline knots          & \multicolumn{1}{c}{--}  & \multicolumn{1}{c}{--}  & 0.1 \\
    Angular efficiency                     	    & 0.5 & 0.1 & \multicolumn{1}{c}{--}  \\
    Efficiency correlation                 	    & 0.2 & 0.4 & 0.6 \\
    Fake tracks from noisy hits 	            & 0.1 & \multicolumn{1}{c}{--}  & 0.3 \\
    $\decay{\Bc}{\Bs\pip}$ contamination        & \multicolumn{1}{c}{--}  & \multicolumn{1}{c}{--}  & 0.5 \\
    \hline
    \textbf{Total} & 7.3 & 7.3 & 3.7 \\
    \hline
    \textbf{Statistical} & 33.8 & 25.7 & 2.2 \\
    \hline
  \end{tabular}
\end{center}
\end{table}
%%%%%%%%%%%%%%%%%%%%%%%%%%%%%%%%%

The uncertainties from the $m(\jpsi\pip\pim)$ fit model
comprise those from both the signal and background models.
The default Hypatia function is replaced by a double-sided Crystal Ball function~\cite{Skwarnicki:1986xj} for the signal, 
and the default fifth-order Chebyshev polynomial function is replaced with a fourth-order Chebyshev polynomial function for the background.
The resulting variations between the alternative and baseline fits are taken as the systematic uncertainties.
These uncertainties are summed in quadrature. 

The uncertainty from the $\pip\pim$ resonance model is evaluated by including the $f_0(500)$, $\rho(770)$ and $f_0(1370)$ resonances, respectively.
The $f_0(500)$ resonance is modelled using the Bugg model~\cite{Bugg:2006gc},
the $\rho(770)$ resonance is described with the Gounaris-Sakurai function~\cite{Gounaris:1968mw},
and the $f_0(1370)$ resonance is modelled with the RBW function.
The largest deviation is assigned as the systematic uncertainty.

The systematic uncertainties related to fixed parameters are independently evaluated 
by varying each fixed parameter within its uncertainties and repeating the final fit. 
This procedure is applied to the decay width of the light mass eigenstate, $\GL$, 
and the mass difference, $\dms$, whose uncertainties are taken from PDG~\cite{PDG2024}.
The systematic uncertainties on the pole masses and decay widths of $f_2(1270)$ and $f^\prime_2(1525)$ resonances are
considered in the same way, using the uncertainties in Table~\ref{tab_infor_resonance}, and are further summed in quadrature.
The production asymmetry, $A_{\rm P}$, has not yet been measured at $\sqrt{s} = 13\tev$, 
therefore a conservative variation of $\pm 2\%$ is taken according to the measurements at $\sqs = 7\tev$ and $8\tev$~\cite{LHCb-PAPER-2016-062}.

The systematic uncertainties due to the decay-time resolution uncertainties and the differences between the control and signal channels
are estimated as follows.
The relative uncertainty of decay-time resolution is less than 1\%,
and a 2\% deviation between the control and signal channels is observed using corresponding simulations.
The width of the resolution model is conservatively constrained to a 5\% variation in the fit, and
the induced systematic uncertainties are calculated as the sum in quadrature of the shift
and the difference in statistical uncertainties between the alternative fit and baseline fit.
Due to vertex detector misalignment, the prompt $\jpsi\pip\pim$ sample shows a decay-time bias in the decay-time resolution study~\cite{LHCb-PAPER-2023-016}.
The final fit is repeated after shifting the mean of the decay-time resolution model by the largest observed time bias in the study of decay-time resolution, and 
the resulting differences in the measured parameters are treated as systematic uncertainties. 

The effective decay-time resolutions are recalculated with the dilution method~\cite{LHCb-PAPER-2023-016},
and the data are refitted with the obtained alternative calibration parameters.
The resulting changes are regarded as systematic uncertainties.
The systematic uncertainties associated with the calibration model of decay-time resolution are obtained
by replacing the linear model with a quadratic model,
and those associated with the predicted decay-time uncertainties binning in the calibration are estimated with an alternative eight-subsample scheme.
Both uncertainties are found to be negligible.

Due to the finite size of the simulation sample,
the systematic uncertainties for the coefficients of decay-time efficiency are studied using pseudoexperiments,
where~250 sets of coefficients are 
generated according to their uncertainties and correlations.
These generated coefficients are used to refit the data. 
The distribution of each physics parameter is fitted with a Gaussian function,
whose width is quoted as the systematic uncertainty.
The sensitivity to the knots of the cubic spline function used to model the decay-time efficiency is studied by an alternative set of knots, and 
the resulting shifts are set as systematic uncertainties.
The systematic uncertainties originating from the coefficients of the angular efficiency are studied in a similar way to those of the decay-time efficiency,
except that two Gaussian functions are used to fit the bimodal distribution of $\phis$ 
and the separation is adopted as the systematic uncertainty.
The influence of the PID probabilities on the correction of the simulation in the angular efficiency determination 
is evaluated 
by performing the determination of the angular efficiency without the PID probabilities,
and the observed shifts are negligible.
A correlation between decay time and $\cospi$
is observed, 
and the associated systematic uncertainties are estimated with a similar pseudoexperiment method,
where~500 pseudoexperiment samples are generated 
with time-dependent angular efficiencies
and each pseudoexperiment sample is fitted with the default angular efficiency.
A Gaussian function is used to fit the distribution of each physics parameter,
and the resulting shift is taken as the systematic uncertainty.

The systematic uncertainties due to the imperfect removal of fake tracks reconstructed from noisy hits are
evaluated by including these simulated events 
in the determination of decay-time and angular efficiencies, and executing the final fit again. 
The induced deviations are taken as systematic uncertainties.
Approximately 0.8\% of the selected signal candidates are expected to originate from $\decay{\Bc}{\Bs\pip}$ decay~\cite{LHCb-PAPER-2014-059},
and this leads to a systematic uncertainty of $0.0005\invps$ 
in~$\GH$~\cite{LHCb-PAPER-2019-003}. 

\section{Conclusion}
\label{sec:Conclusion}

A time-dependent flavour-tagged amplitude analysis of the $\decay{\Bs}{\jpsi\pip\pim}$ decays is performed 
using the \lhcb full \runtwo data sample corresponding to an integrated luminosity of $6\invfb$. 
A data sample of approximately 94\,000 $\decay{\Bs}{\jpsi\pip\pim}$ signal decays, 
with an averaged decay-time resolution of $38\fs$
and an averaged tagging power of $4.5\%$
is used to measure the \CP-violation parameters to be 
\begin{equation*}
    \begin{aligned}
 \phis    &= -0.077 \pm 0.034 \pm 0.007\rad,\\
 \abs{\lambda} &= 0.993 \pm 0.026 \pm 0.007,\\
 \GH   &= 0.610 \pm 0.002 \pm 0.004\invps,
    \end{aligned}
\end{equation*}
where the first uncertainty is statistical and the second systematic. 
This measurement is consistent with the previous \lhcb measurements
~\cite{LHCb-PAPER-2014-059,LHCb-PAPER-2023-016,LHCb-PAPER-2017-008,LHCb-PAPER-2020-042,LHCb-PAPER-2014-019,LHCb-PAPER-2019-003,LHCb-PAPER-2016-027,LHCb-PAPER-2014-051,LHCb-PAPER-2024-027}
and supersedes the measured values in Ref.~\cite{LHCb-PAPER-2019-003}.
No evidence of \CP violation is observed in the $\decay{\Bs}{\jpsi\pip\pim}$ decays.

The measurements are combined with the previous results from the \runone data sample~\cite{LHCb-PAPER-2014-019,LHCb-PAPER-2019-003}.
A correlation of 100\% is assumed for shared systematic sources, 
and 0\% for the remaining ones.
The combination yields
$\phis    = -0.046 \pm 0.031\rad$,
$\abs{\lambda} = 0.975 \pm 0.024$,
and $\GH   = 0.610 \pm 0.004\invps$.
The uncertainty includes the statistical and systematic uncertainties.
This combination gives $\chisqndf = 7/2$.
The correlations between the parameters are 
$\rho(\phis,\,\abs{\lambda}) = 0.059$, $\rho(\abs{\lambda},\,\GH) = 0.010$, and $\rho(\phis,\,\GH) = 0.016$.
%%%%%%%%%%%%%%%%%%%%%%%%%%%%%%%%%%%%%%%%%%
\begin{figure}[!tb]
    \begin{center}
        \includegraphics[width=0.9\linewidth]{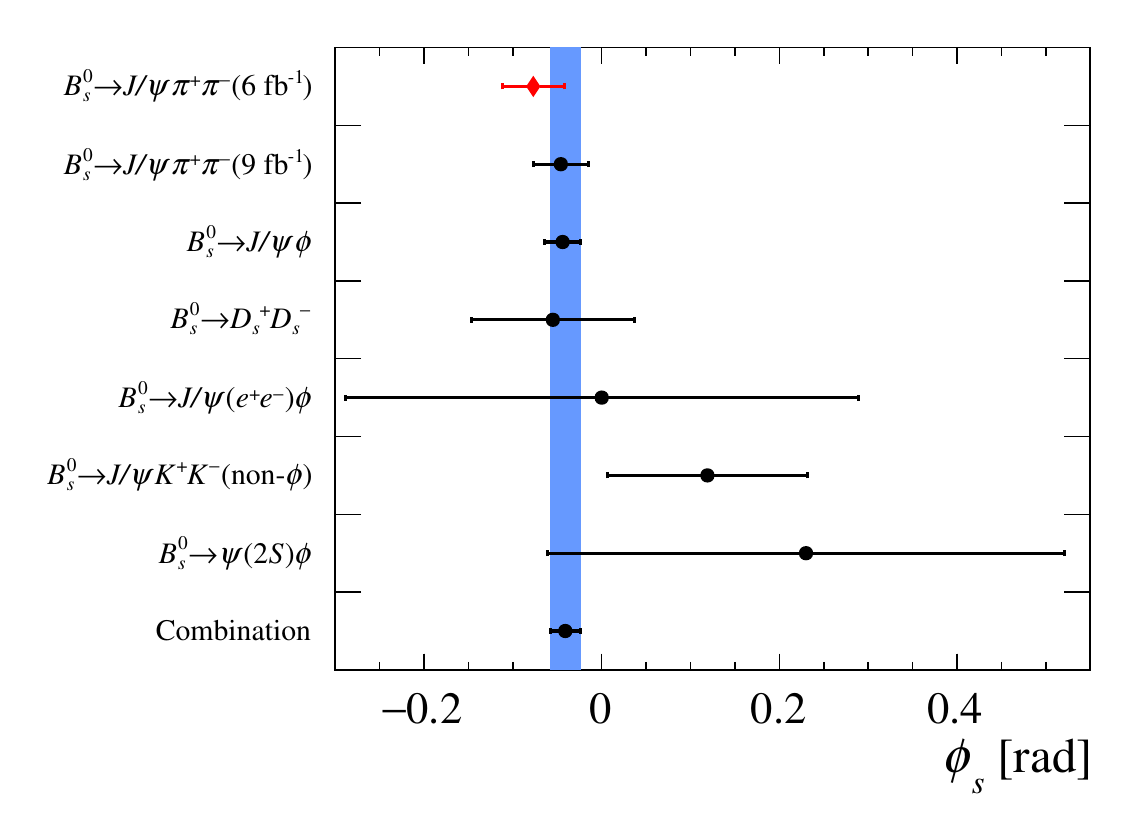}
        \vspace*{-0.5cm}
    \end{center}
    \caption{
    The $\mbox{LHCb}\xspace$ measurements of $\phi_s$ for the decays of $B_{s}^{0} \!\to J/\psi \pi^{+}\pi^{-}$ ($6\ensuremath{\fb^{-1}}\xspace$, this work), $B_{s}^{0} \!\to J/\psi \pi^{+}\pi^{-}$ ($9\ensuremath{\fb^{-1}}\xspace$, this work and $\mbox{Run 1}\xspace$),
    $B_{s}^{0} \!\to J/\psi \phi$~\cite{LHCb-PAPER-2023-016}, $B_{s}^{0} \!\to D_{s}^{+}D_{s}^{-}$~\cite{LHCb-PAPER-2024-027}, 
    $B_{s}^{0} \!\to J/\psi(e^{+}e^{-}) \phi$~\cite{LHCb-PAPER-2020-042}, $B_{s}^{0} \!\to J/\psi K^{+}K^{-}$ (with \mbox{$m_{KK}>1.05\ensuremath{\aunit{Ge\kern -0.1em V\!/}c^2}\xspace$})~\cite{LHCb-PAPER-2017-008},
    $B_{s}^{0} \!\to \psi{(2S)} \phi$~\cite{LHCb-PAPER-2016-027}, 
    along with the combination of all measurements, are presented.
    Unless stated otherwise, $J/\psi$ and $\psi(2S)$ decay to $\mu^{+}\mu^{-}$.
    }
    \label{Fig5}
\end{figure}
%%%%%%%%%%%%%%%%%%%%%%%%%%%%%%%%%%%%%%%%%%
Figure~\ref{Fig5} shows the measurements of $\phis$ performed by the \lhcb experiment in different decay channels.
A further combination of all \lhcb measurements~\cite{LHCb-PAPER-2014-059,LHCb-PAPER-2023-016,LHCb-PAPER-2017-008,LHCb-PAPER-2020-042,LHCb-PAPER-2014-019,LHCb-PAPER-2016-027,LHCb-PAPER-2014-051,LHCb-PAPER-2024-027} 
results in $\phis = -0.041 \pm 0.017\rad$ with $\chisqndf=35/30$.
This value is consistent with the average in the $\decay{\Bs}{\jpsi\phi}$ decay~\cite{LHCb-PAPER-2023-016} 
and improves the precision by approximately $15\%$.

% Do not include this in any draft (just for information in the template)
%\input{LHCbInternal/acknowledgements_template}
% Comment this in for paper drafts; do not include this in analysis note, conference and figure reports
\section*{Acknowledgements}
%
% These Acknowledgements valid from 16/06/2026
%
\noindent We express our gratitude to our colleagues in the CERN
accelerator departments for the excellent performance of the LHC. We
thank the technical and administrative staff at the LHCb
institutes.
We acknowledge support from CERN and from the national agencies:
ARC (Australia);
CAPES, CNPq, FAPERJ and FINEP (Brazil); 
MOST and NSFC (China); 
CNRS/IN2P3 and CEA (France);  % added CEA 26/02/2026
BMFTR, DFG and MPG (Germany);
NKFIH (Hungary);              % added 16/06/2026
INFN (Italy); 
NWO (Netherlands); 
MNiSW and NCN (Poland); 
MEC/IFA (Romania); 
%MSHE (Russia); 
MICIU and AEI (Spain);
SNSF and SER (Switzerland); 
NASU (Ukraine); 
STFC (United Kingdom); 
DOE NP and NSF (USA).
%%%%%%%%%%%%%%%%%%%%%%%%%%%%%%%%%%%%%%%%%%%%%
We acknowledge the computing resources that are provided by ARDC (Australia), 
CBPF (Brazil),
CERN, 
IHEP and LZU (China),
IN2P3 (France), 
KIT and DESY (Germany), 
INFN (Italy), 
SURF (Netherlands),
Polish WLCG (Poland),
IFIN-HH (Romania), % http://dx.doi.org/10.13039/100019931,"Institutul National de Cercetare-Dezvoltare pentru Fizica si Inginerie Nucleara 'Horia Hulubei'"
%RRCKI and Yandex LLC (Russia), 
PIC (Spain), CSCS (Switzerland), 
GridPP (United Kingdom),
and NSF (USA).  % added Feb2026
%%%%%%%%%%%%%%%%%%%%%%%%%%%%%%%%%%%%%%%%%% 
We are indebted to the communities behind the multiple open-source
software packages on which we depend.
%%%%%%%%%%%%%%%%%%%%%%%%%%%%%%%%%%%%%%%%%%
Individual groups or members have received support from
% ARC and ARDC (Australia); % moved to national 16/01/2025
RTP (Australia), % added 06/03/2026
FWO Odysseus grant G0ASD25N (Belgium), % added 20/4/2026
Key Research Program of Frontier Sciences of CAS, CAS PIFI, CAS CCEPP, Fundamental Research Funds for the Central Universities (China); 
%and Sci.\
%\& Tech.\ Program of Guangzhou (China); Removed 24/11/25
Minciencias (Colombia);
EPLANET, Marie Sk\l{}odowska-Curie Actions, ERC and NextGenerationEU (European Union);
A*MIDEX, ANR, IPhU and Labex P2IO, and R\'{e}gion Auvergne-Rh\^{o}ne-Alpes (France);
%RFBR, RSF and Yandex LLC (Russia);
Alexander-von-Humboldt Foundation (Germany);
ICSC (Italy); 
%GVA, XuntaGal, GENCAT, Inditex, InTalent and Prog.~Atracci\'on Talento, CM (Spain);
Severo Ochoa and Mar\'ia de Maeztu Units of Excellence, GVA, XuntaGal, GENCAT, InTalent-Inditex and Prog.~Atracci\'on Talento CM (Spain);
%XuntaGal --> Xunta de Galicia 
% SRC (Sweden);  % removed 27/02/2026 - end of grant
the Leverhulme Trust, the Royal Society and UKRI (United Kingdom).

%\input{LHCbInternal/supplementary_template}

%\input{appendix}

% This should be taken out in the final paper
%\input{supplementary}

\addcontentsline{toc}{section}{References}
%\setboolean{inbibliography}{true}
\bibliographystyle{LHCb/LHCb}
\bibliography{main,LHCb/standard,LHCb/LHCb-PAPER,LHCb/LHCb-CONF,LHCb/LHCb-DP,LHCb/LHCb-TDR}

\newpage
% LHCb collaboration author list
% Data extracted on July 20th, 2026 at 1:57pm for paper reference LHCb-PAPER-2026-017
\centerline
{\large\bf LHCb collaboration}
\begin
{flushleft}
\small
R.~Aaij$^{39}$\lhcborcid{0000-0003-0533-1952},
M.~Abdelfatah$^{71}$,
A.S.W.~Abdelmotteleb$^{59}$\lhcborcid{0000-0001-7905-0542},
C.~Abellan~Beteta$^{53}$\lhcborcid{0009-0009-0869-6798},
F.~Abudin\'en$^{61}$\lhcborcid{0000-0002-6737-3528},
T.~Ackernley$^{63}$\lhcborcid{0000-0002-5951-3498},
A.A.~Adefisoye$^{71}$\lhcborcid{0000-0003-2448-1550},
B.~Adeva$^{49}$\lhcborcid{0000-0001-9756-3712},
M.~Adinolfi$^{57}$\lhcborcid{0000-0002-1326-1264},
P.~Adlarson$^{87,44}$\lhcborcid{0000-0001-6280-3851}
% ...

{\footnotesize \it

% ...
$^{39}$Nikhef National Institute for Subatomic Physics, Amsterdam, Netherlands\\
% ...
$^{44}$National Center for Nuclear Research (NCBJ), Warsaw, Poland\\
% ...
$^{49}$Instituto Galego de F{\'\i}sica de Altas Enerx{\'\i}as (IGFAE), Universidade de Santiago de Compostela, Santiago de Compostela, Spain\\
% ...
$^{53}$Physik-Institut, Universit{\"a}t Z{\"u}rich, Z{\"u}rich, Switzerland\\
% ...
$^{57}$H.H. Wills Physics Laboratory, University of Bristol, Bristol, United Kingdom\\
% ...
$^{59}$Department of Physics, University of Warwick, Coventry, United Kingdom\\
% ...
$^{61}$School of Physics and Astronomy, University of Edinburgh, Edinburgh, United Kingdom\\
$^{62}$School of Physics and Astronomy, University of Glasgow, Glasgow, United Kingdom\\
$^{63}$Oliver Lodge Laboratory, University of Liverpool, Liverpool, United Kingdom\\
% ...
$^{71}$Syracuse University, Syracuse, NY, United States\\
% ...
$^{87}$Department of Physics and Astronomy, Uppsala University, Uppsala, Sweden, associated to $^{62}$\\
% ...
}
\end{flushleft}
%\input{~/Downloads/Authorship_LHCb-SD-2024-001-grouped-internal-use-only-do-not-publish-2.tex}

%The author list for journal publications is generated from the
%Membership Database shortly after 'approval to go to paper' has been
%given.  It is available at \url{https://lbfence.cern.ch/membership/authorship}
%and will be sent to you by email shortly after a paper number
%has been assigned.  
%The author list should be included in the draft used for 
%first and second circulation, to allow new members of the collaboration to verify
%that they have been included correctly. Occasionally a misspelled
%name is corrected, or associated institutions become full members.
%Therefore an updated author list will be sent to you after the final
%EB review of the paper.  In case line numbering doesn't work well
%after including the authorlist, try moving the \verb!\bigskip! after
%the last author to a separate line.
%
%
%The authorship for Conference Reports should be ``The LHCb
%collaboration'', with a footnote giving the name(s) of the contact
%author(s), but without the full list of collaboration names.
%
%
%The authorship for Figure Reports should be ``The LHCb
%collaboration'', with no contact author and without the full list 
%of collaboration names.

\end{document}